\documentclass[aps,pre,reprint,amsmath,amssymb,showpacs,superscriptaddress]{revtex4-1}
\usepackage[right]{lineno}
\usepackage{graphicx}
\usepackage{dcolumn}
\usepackage{bm}
\usepackage{braket}%
\usepackage{natbib}
\usepackage{hyperref}
\hypersetup{%
 colorlinks=true, citecolor=blue, filecolor=blue, linkcolor=blue, urlcolor=blue}

\begin{document}

\preprint{Manuscript}

\title{Color-gradient lattice Boltzmann model with nonorthogonal central moments: \\ Hydrodynamic melt-jet breakup simulations}



\author{Shimpei Saito}
\email[]{saito.shimpei@gmail.com}
\affiliation{University of Tsukuba, Tsukuba 305-8573, Japan}

\author{Alessandro De Rosis}
\altaffiliation{Present address: Calle de Toledo 77, 28005, Madrid, Spain}\affiliation{Technion -- Israel Institute of Technology, Haifa 32000, Israel}

\author{Alessio Festuccia}
\affiliation{University Niccol{\`{o}} Cusano, 00166 Rome, Italy}

\author{Akiko Kaneko}
\affiliation{University of Tsukuba, Tsukuba 305-8573, Japan}

\author{Yutaka Abe}
\affiliation{University of Tsukuba, Tsukuba 305-8573, Japan}

\author{Kazuya Koyama}
\affiliation{Mitsubishi FBR Systems, Inc., Tokyo 150-0001, Japan}

\date{\today}

\begin{abstract}
	We develop a lattice Boltzmann (LB) model for immiscible two-phase flow simulations with central moments (CMs).
	This successfully combines a three-dimensional nonorthogonal CM-based LB scheme [A. De Rosis, \href{http://doi.org/10.1103/PhysRevE.95.013310}{Phys. Rev. E {\bf 95}, 013310 (2017)}] with our previous color-gradient LB model [S. Saito, Y. Abe, and K. Koyama, \href{http://doi.org/10.1103/PhysRevE.96.013317}{Phys. Rev. E {\bf 96}, 013317 (2017)}].
	Hydrodynamic melt-jet breakup simulations show that the proposed model is significantly more stable, even for flow with extremely high Reynolds numbers, up to $O(10^6)$. 
	This enables us to investigate the phenomena expected under actual reactor conditions.
\end{abstract}


\maketitle

\section{Introduction \label{sec:intro}}

	Multiphase and multicomponent flows appear in many natural and industrial processes.
	A liquid jet injected into another fluid is an interesting example of such a flow, 
	and understanding the breakup of liquid jets has been a topic of significant interest for more than a century.
	Since the pioneering works of \citet{Plateau1873} and \citet{Rayleigh1878}, this subject has been extensively studied both theoretically, experimentally, and numerically~\citep{McCarthy1974,Lin1998,Villermaux2007,Eggers2008}.
		In the linear theory framework, the liquid jet breakup problem is described in terms of the density ratio $\gamma$, viscosity ratio $\eta$, Reynolds number $\mathrm{Re}$, Weber number $\mathrm{We}$, and Froude number $\mathrm{Fr}$ as follows~\citep{Lin1998}:
\begin{align}
	\gamma =& \frac{\rho_j}{\rho_c}, \label{eq:densityRatio}
	\\
	\eta =& \frac{\nu_j}{\nu_c}, \label{eq:viscosityRatio}
	\\
	\mathrm{Re} =& \frac{\rho_j  u_{j0} D_{j0}}{\mu_j}, \label{eq:jetReynolds}
	\\
	\mathrm{We} =& \frac{\rho_j u_{j0}^2 D_{j0}}{\sigma}, \label{eq:jetWeber}
	\\
	\mathrm{Fr} =& \frac{u_{j0}^2}{gD_{j0}}, \label{eq:Froude}
\end{align}
where $\rho$ is the density, $\nu$ is the kinematic viscosity, $u_{j0}$ is the jet velocity, $D_{j0}$ is the jet inlet diameter, $\sigma$ is the interfacial tension, and $g$ is the acceleration due to gravity. 
	The subscript $j$ and $c$ denote the dispersed and the continuous phases, respectively.

	At low injection velocities, drops form directly at the nozzle, while at higher velocities a liquid jet issues from the nozzle and then breaks into various droplet patterns.
	Discovering when these regimes occur is of significant interest in the study of liquid-jet breakup.
	\citet{Ohnesorge1936} classified his results into four breakup regimes: dripping (0), varicose (I), sinuous (II), and atomization (III)~\citep{Kolev2005,McKinley2011}.
	He also provided a map of these regimes for liquid jets in a gas in terms of the Reynolds number $\mathrm{Re}$ and the Ohnesorge number $\mathrm{Oh}$, where
	$\mathrm{Oh}= \mathrm{We}^{1/2}/\mathrm{Re}$ and can thus be calculated using Eqs.~(\ref{eq:jetReynolds}) and (\ref{eq:jetWeber}).
	Following Ohnesorge's work, there has been much research on this subject (see, e.g., \citep{Merrington1947,Tanasawa1954,Grant1966}),
	most of which has focused on liquid--gas systems (liquid jets into gaseous atmospheres). 	Jet breakup in liquid--liquid systems (liquid jets into other liquids) has not been investigated as extensively.

\begin{figure*}[tb]
\begin{center}
	\includegraphics[width=13.5cm]{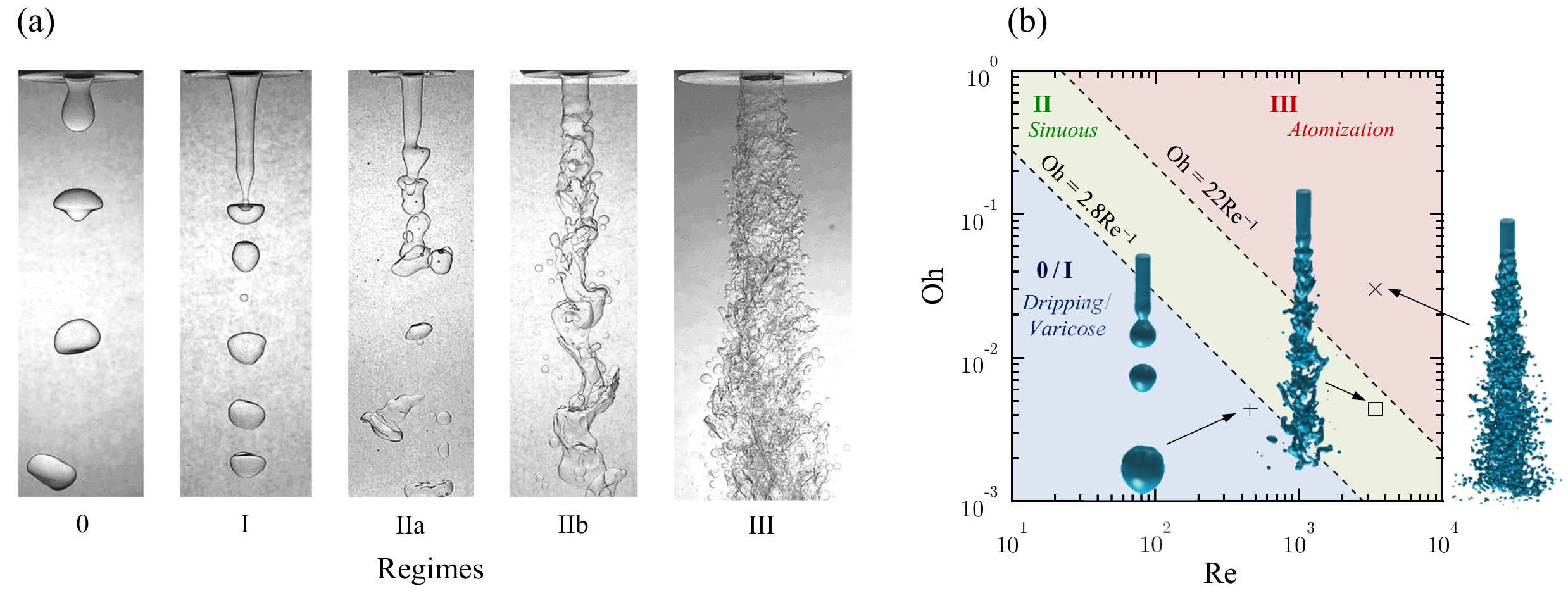}
	\caption{(a) Snapshots of typical jet breakup regimes in liquid--liquid systems: dripping (0), varicose (I), sinuous {\it without} entrainment (IIa), sinuous {\it with} entrainment (IIb), and atomization (III)~\citep{Saito2017}.
	(b) Map of jet breakup regimes in immiscible liquid--liquid systems~\citep{Saito2017,Saito2017b}
	\label{fig:typicalRegimes}}
\end{center}
\end{figure*}

	Recently, \citet{Saito2017} used a series of observations to classify the jet breakup regimes in liquid--liquid systems, as shown in Fig.~\ref{fig:typicalRegimes}(a),
	extending Ohnesorge's classification scheme for liquid--gas systems~\citep{Ohnesorge1936,Kolev2005,McKinley2011}.
	This classification largely follows Ohnesorge's one, but it further divides Regime II into two sub-regimes: sinuous {\it without} entrainment (IIa) and sinuous {\it with} entrainment (IIb).
	On the basis of observations and phenomenological considerations, they derived the following flow-transition criteria~\citep{Saito2017}:
\begin{equation}
	{\rm Oh} = 2.8 {\rm Re}^{-1}, 
\end{equation}
for Regimes I and II, and
\begin{equation}
	{\rm Oh} = 22 {\rm Re}^{-1}, 
\end{equation}
for Regimes II and III. 
	These criteria can be used to predict the breakup regimes of immiscible liquid--liquid jets based on their initial parameters.
	We have successfully reproduced these flow-transition criteria in numerical simulations~\citep{Saito2017b} based on their multiple-relaxation-time (MRT) lattice Boltzmann (LB) two-phase flow model, as illustrated in Fig.~\ref{fig:typicalRegimes}(b).
	
	Liquid--liquid-jet systems can be found in several fields, e.g., chemical processing~\citep{Meister1969,Takahashi1971,Das1997} and CO$_2$ storage in oceans~\citep{Riestenberg2004,Tsouris2007}.
	In the nuclear engineering field, it is important to fully understand the interactions between melt and coolant when designing nuclear reactor safety.
	As a result, the dispersion of liquid metal in water has been extensively investigated in the literature~\citep{Kondo1995,Dinh1999,Abe2006,Matsuo2013},
	going all the way back to G.I. Taylor's classic experiment with mercury and water~\citep{Taylor1963}.
	In these experiments, high-temperature melt and water are often used to simulate the core melt materials and coolant.
	Using the experimental facility at University of Tsukuba (UT), Matsuo {\it et al.}~\citep{Matsuo2008,Matsuo2013} injected molten alloy with a melting point of 78 $\rm{^oC}$ (U-Alloy78) into a water pool, using high-speed visualization to help understand the mechanism behind melt-jet breakup in water.
	The experiments of \citet{Magallon1992}, called FARO-TERMOS (FT), are unique in that they used a liquid sodium coolant: 
	they poured pure molten UO$_2$ into a pool of liquid sodium.
	The fragment size analysis showed that fine fragments were generated by interactions between the molten UO$_2$ and the liquid sodium, but they obtained little physical insight into the breakup behavior because liquid sodium is not transparent.
	In this paper, our simulation targets are these two melt-coolant experiments~\citep{Matsuo2013,Magallon1992}; we call them the UT and FT experiments, respectively.

\begin{figure}[tb]
\begin{center}
	\includegraphics[width=8.5cm]{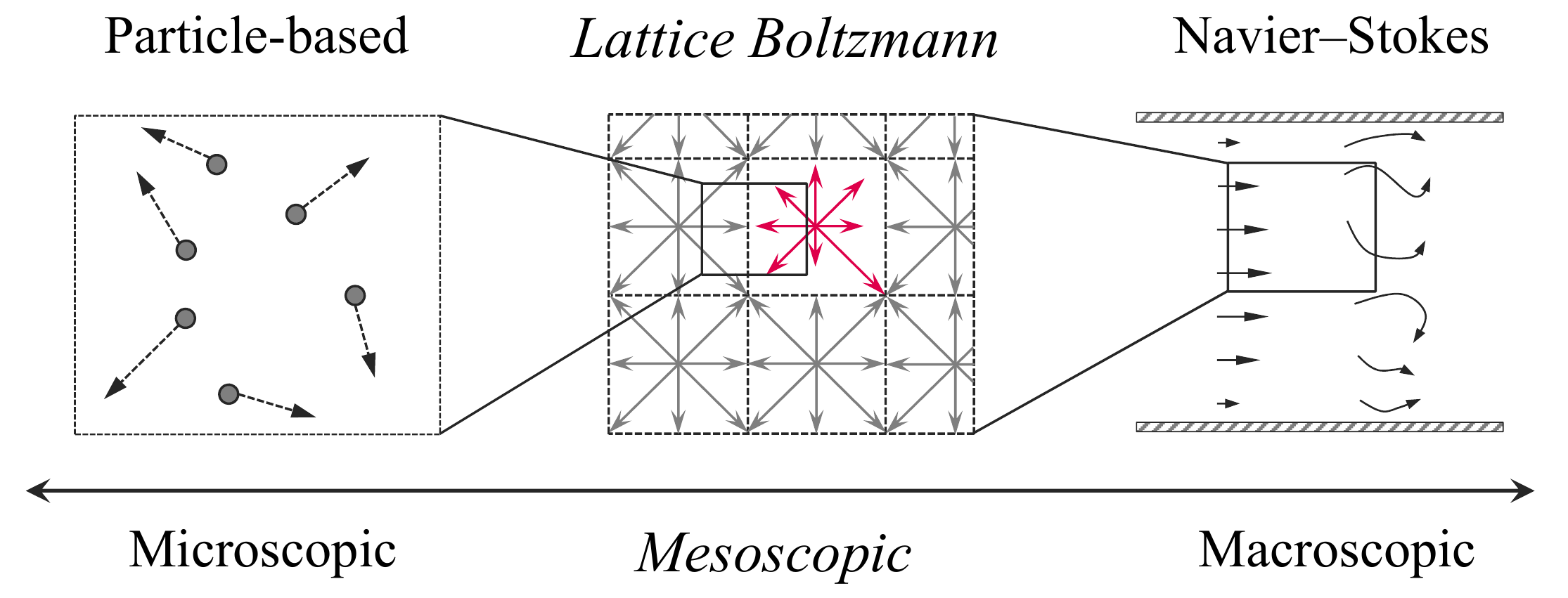}
	\caption{Fluid flow properties revealed at different scales by different simulation methods.
	The LB method is a mesoscopic simulation approach that lies between microscopic particle-based (e.g., molecular dynamics) and macroscopic Navier--Stokes-based methods.
	\label{fig:scale}}
\end{center}
\end{figure}

 The complexity of the phenomena involved in melt-coolant interactions means it is difficult to understand all the mechanisms simultaneously. 
	Investigating the hydrodynamic interactions separately will thus help us to better understand the fundamental melt-jet breakup processes.
	Several researchers have already attempted to simulate jet breakup behavior using the volume-of-fluid method (see, e.g., \citep{Thakre2015,Secareanu2016,Zhou2017}).
	In this paper, we use the LB method for multiphase flows, which has come to be recognized as a powerful tool for analyzing complex fluid dynamics, including multicomponent and multiphase flows~\citep{Aidun2010}.
	Figure~\ref{fig:scale} illustrates the fluid flow properties at different scales. 
	Compared with macroscopic CFD methods, which are based on the Navier--Stokes equations, the LB method, which uses mesoscopic kinetic equations, has several advantages,
	such as making it easy to incorporate mesoscale physics like interfacial breakup and coalescence.
	In addition, the computational cost of simulating realistic fluid flows is far more reasonable than with particle-based methods (e.g., molecular dynamics).

	Two-phase or multiphase LB models can be divided into four categories, namely, 
	color-gradient~\cite{Gunstensen1991,Grunau1993}, pseudopotential~\cite{Shan1993,Shan1994}, free-energy~\cite{Swift1995,Swift1996}, and mean-field~\cite{He1999} models.
	This is not an exhaustive classification; for instance, the latter two model types are sometimes called phase-field models~\citep{Li2016} since the Cahn--Hilliard (or similar) interface tracking equations can be derived from them.
	For further details about multiphase LB models, interested readers can refer to several comprehensive review papers~\citep{Chen1998,Nourgaliev2003,Aidun2010,Chen2014,Liu2015,Li2016} and references therein.
	This paper focus on color-gradient (CG) models, as they have many strengths for simulating multiphase or multicomponent flows, including strict mass conservation for each fluid and flexibility in adjusting the interfacial tension~\citep{Ba2016}. 
	They also do not require us to use the static drop test to determine the interfacial tension, as this can be obtained directly without further analysis or assumptions. 
	In addition, CG models exhibit very low dissolution for tiny droplets or bubbles~\citep{Liu2015}.

	CG models, often called R-K models, were first developed by \citet{Gunstensen1991}, who extended Rothman and Keller's two-component lattice gas automata model~\citep{Rothman1988}. 
	Later, \citet{Grunau1993} enabled density and viscosity ratios to be introduced by modifying the forms of the distribution functions. 
	\citet{Latva-Kokko2005} then replaced Gunstensen's maximization-recoloring step with a formulaic segregation algorithm.
	Instead of widening the interface, Latva-Kokko--Rothman's recoloring algorithm solves two issues with the previous CG models, namely, the lattice-pinning problem and spurious velocities.
	\citet{Reis2007} extended the model to a common two-dimensional nine-velocity (D2Q9) lattice and
	modified the perturbation operator to correctly recover the Navier--Stokes equations.
	\citet{Leclaire2012} demonstrated that combining Latva-Kokko--Rothman's recoloring operator~\citep{Latva-Kokko2005} with Reis--Phillips' perturbation operator~\citep{Reis2007} greatly improves the numerical stability and accuracy of the solutions over a wide parameter range. 
	Using an isotropic gradient operator also enhances the numerical stability and accuracy~\citep{Leclaire2011}.
	\citet{Liu2012} derived a generalized perturbation operator using the phase-field (or order parameter), and formulated the CG model in three dimensions.
	\citet{Leclaire2017} generalized the CG model to two and three dimensions.
	For interested readers, Leclaire's MATLAB scripts will help to understand how to code the CG model~\citep{Leclaire2013c}


The so-called BGK~\citep{Bhatnagar1954} approximation refers to this simplest form of the collision operator, which forces all populations to relax towards an equilibrium state with the same rate. Despite its simplicity and phenomenal popularity, the BGK LB method is known to suffer from numerical instability under high-Re (low-viscosity) conditions. 
	One way to overcome this issue is to modify the collision operator~\citep{Luo2011}.
	For example, MRT collision operators~\citep{dHumieres1994,Lallemand2000,dHumieres2002} have been widely used, even for multiphase flows, to enhance numerical stability and accuracy and reduce spurious currents near the interface.	
	Later, \citet{Geier2006} proposed a new collision operator based on the relaxation of central moments (CMs), that can be obtained by shifting the lattice directions according to the local fluid velocity.
	Many studies have developed this approach to fully exploit the properties of CM-based schemes (see, e.g., De Rosis~\citep{DeRosis2017d} and references therein).
	For multiphase flows, \citet{Lycett-Brown2014} first introduced CMs into the pseudopotential multiphase LB model.
	\citet{Leclaire2014b} also introduced CMs into the CG model with unit density ratio.
	Very recently, \citet{DeRosis2018} formulated a CM-based LB scheme for coupled Cahn--Hilliard--Navier--Stokes equations.
	De Rosis has consistently adopted nonorthogonal CMs~\citep{DeRosis2016,DeRosis2017a,DeRosis2017b,DeRosis2017c, DeRosis2017e}, which are characterized by straightforward derivation and easy practical implementation. 
	Moreover, his analytical formulation is very general, as it can be extended to any lattice velocity space.

	In this paper, we present a three-dimensional CG LB model and apply it to hydrodynamic simulations of melt-jet breakup.
	Sec.~\ref{sec:methodology} describes the formulation of this LB model.
	Sec.~\ref{sec:laplace} uses numerical tests on static droplets to evaluate the proposed model. 
	Sec.~\ref{sec:jetBreakup} applies the model to simulating melt-jet breakup under the conditions of the $UT$ and {\it FT} experiments.
	Finally, Sec.~\ref{sec:conclusions} concludes this paper.

\section{Methodology\label{sec:methodology}}
	The LB model presented here is based on our previous MRT CG model~\citep{Saito2017b}.
	The most significant difference between the current and previous models is the introduction of De Rosis' nonorthogonal CMs~\citep{DeRosis2016,DeRosis2017a}.
	In the current three-dimensional LB model, the distribution functions move on a three-dimensional 27-velocity (D3Q27) lattice~\cite{Succi2001}. 
	We adopt a lattice speed $c=\delta_x/\delta_t=1$, where $x$ and $t$ are the
lattice spacing and time step, respectively.
	The lattice velocities ${\bf c}_i = [\ket{c_{ix}},\ket{c_{iy}},\ket{c_{iz}}]$ are defined as follows:
\begin{equation}
	\begin{split}
	\ket{c_{ix}} = &[0,  1,-1, 0, 0, 0, 0,  1,-1, 1,-1, 0, 0, 0, 0, 1,-1, 1,-1, \\
	& 1,-1, 1,-1, 1,-1,-1, 1]^\top, \\
	\ket{c_{iy}} = &[0,  0, 0, 1,-1, 0, 0,  1,-1,-1, 1, 1,-1, 1,-1, 0, 0, 0, 0,  \\
	& 1,-1, 1,-1,-1, 1, 1,-1]^\top, \\
	\ket{c_{iz}} = &[0,  0, 0, 0, 0, 1,-1,  0, 0, 0, 0, 1,-1,-1, 1, 1,-1,-1, 1,  \\
	& 1,-1,-1, 1, 1,-1, 1,-1]^\top, \label{eq:velVec}
	\end{split}
\end{equation}
where $i~(=0,1,\dots,26)$ represents the lattice-velocity directions and the superscript ``$\top$'' is the transpose operator.
	Here, we employ Dirac's bracket notation, where the ``bra'' operator $\bra{\cdot}$ denotes a row vector along one of the lattice-velocity directions and the ``ket'' operator $\ket{\cdot}$ denotes a column vector.

	The model represents two immiscible fluids as red and blue fluids. 
	Distribution functions $\displaystyle f_i^k$ represent the fluids $k$, 
where $k=r$ and $b$ denote ``red'' and ``blue,'' respectively,
and $i$ is the lattice-velocity direction.
	The total distribution function is defined as $f_i = f_i^r+f_i^b$,
	and the evolution is expressed by the following LB equation:
\begin{equation}
	f_i^k({\bf x}+{\bf c}_i \delta_t,t+\delta_t) - f_i^k({\bf x},t) = \Omega_i^k({\bf x},t), \label{LBE}
\end{equation}
where ${\bf x} = [x,y,z]$ and $t$ are the position and time, respectively.
	The collision operator $\displaystyle \Omega_i^k$ is made up of three sub-operators~\cite{Tolke2002}: 
\begin{equation}
	\Omega_i^k = (\Omega_i^k )^{(3)} \left[(\Omega_i^k )^{(1)} + (\Omega_i^k)^{(2)} \right],
\end{equation}
where $(\Omega_i^k)^{(1)}$, $(\Omega_i^k)^{(2)}$, and $(\Omega_i^k)^{(3)}$ are the single-phase collision, perturbation, and recoloring operators, respectively.	
	As in Ref.~\cite{Leclaire2017}, the single-phase and perturbation operators are applied using the color blind distribution function $f_i$.

	In this paper, we adopt the general MRT (GMRT) framework~\citep{Fei2017, Fei2018b} to describe the single-phase collision operator with nonorthogonal CMs, due to the simplicity of its relationship to the MRT and SRT collision operators. It should be noted that \citet{Fei2018b} propose a simplified version of De Rosis' nonorthogonal CMs~\citep{DeRosis2016,DeRosis2017a}, showing a significantly reduced computational cost.
	In the GMRT framework, the single-phase collision operator can be written as
\begin{equation}
	\left(\Ket{\Omega}\right)^{(1)} = -{\bf M}^{-1}{\bf N}^{-1}{\bf K}{\bf N}{\bf M}
	\left(\Ket{f}-\ket{f^{(e)}} \right) + \Ket{F}, \label{eq:scoll}
\end{equation}
where ${\bf M}$, ${\bf N}$, and ${\bf K}$ are the transformation, shift~\citep{Asinari2008,Fei2017, Fei2018b}, and relaxation matrices, respectively.
The density of the fluid $k$ is given by
\begin{equation}
	\rho_k = \sum_i f_i^k.
\end{equation}
The total fluid density is given by $\rho = \sum_k \rho_k$, and the total momentum is defined as
\begin{equation}
	\rho{\bf u} = \sum_i f_i {\bf c}_i + \frac{1}{2} {\bf F}\delta_t, \label{Momentum}
\end{equation}
where ${\bf F}$ is the body force.
	Note that, in Eq.~(\ref{Momentum}), the local velocity has been modified to incorporate the spatially varying body force~\citep{Guo2002}.
	To model the single-phase collision operator [Eq.~(\ref{eq:scoll})], we use the nonorthogonal CMs proposed by De Rosis~\citep{DeRosis2017a}, namely,
\begin{equation}
{\bf NM} = {\bf T} = 
\left [
	\begin{array}{c}
		\bra{|{\bf c}_i|^0} \\	
		\bra{\bar{c}_{ix}} \\	
		\bra{\bar{c}_{iy}} \\	
		\bra{\bar{c}_{iz}} \\	
		\bra{\bar{c}_{ix} \bar{c}_{iy}} \\	
		\bra{ \bar{c}_{ix} \bar{c}_{iz} } \\ 
		\bra{ \bar{c}_{iy} \bar{c}_{iz} } \\ 
		\bra{ \bar{c}_{ix}^2 - \bar{c}_{iy}^2 } \\ 
		\bra{ \bar{c}_{ix}^2 - \bar{c}_{iz}^2 } \\ 
		\bra{ \bar{c}_{ix}^2 + \bar{c}_{iy}^2 + \bar{c}_{iz}^2 } \\ 
		\bra{ \bar{c}_{ix}\bar{c}_{iy}^2 + \bar{c}_{ix}\bar{c}_{iz}^2 } \\ 
		\bra{ \bar{c}_{ix}^2\bar{c}_{iy} + \bar{c}_{iy}\bar{c}_{iz}^2 } \\ 
		\bra{ \bar{c}_{ix}^2\bar{c}_{iz} + \bar{c}_{iy}^2\bar{c}_{iz} } \\ 
		\bra{ \bar{c}_{ix}\bar{c}_{iy}^2 - \bar{c}_{ix}\bar{c}_{iz}^2 } \\ 
		\bra{ \bar{c}_{ix}^2\bar{c}_{iy} - \bar{c}_{iy}\bar{c}_{iz}^2 } \\ 
		\bra{ \bar{c}_{ix}^2\bar{c}_{iz} - \bar{c}_{iy}^2\bar{c}_{iz} } \\ 
		\bra{ \bar{c}_{ix}\bar{c}_{iy}\bar{c}_{iz} } \\ 
		\bra{ \bar{c}_{ix}^2\bar{c}_{iy}^2 + \bar{c}_{ix}^2\bar{c}_{iz}^2 + \bar{c}_{iy}^2\bar{c}_{iz}^2 } \\ 
		\bra{ \bar{c}_{ix}^2\bar{c}_{iy}^2 + \bar{c}_{ix}^2\bar{c}_{iz}^2 - \bar{c}_{iy}^2\bar{c}_{iz}^2 } \\ 
		\bra{ \bar{c}_{ix}^2\bar{c}_{iy}^2 - \bar{c}_{ix}^2\bar{c}_{iz}^2 } \\ 
		\bra{ \bar{c}_{ix}^2\bar{c}_{iy}\bar{c}_{iz} } \\ 
		\bra{ \bar{c}_{ix}\bar{c}_{iy}^2\bar{c}_{iz} } \\ 
		\bra{ \bar{c}_{ix}\bar{c}_{iy}\bar{c}_{iz}^2 } \\ 
		\bra{ \bar{c}_{ix}\bar{c}_{iy}^2\bar{c}_{iz}^2 } \\ 
		\bra{ \bar{c}_{ix}^2\bar{c}_{iy}\bar{c}_{iz}^2 } \\ 
		\bra{ \bar{c}_{ix}^2\bar{c}_{iy}^2\bar{c}_{iz} } \\ 
		\bra{ \bar{c}_{ix}^2\bar{c}_{iy}^2\bar{c}_{iz}^2 } 
		\label{eq:momentSet}
	\end{array}
	\right] ,
\end{equation}
where  $\ket{\bar{c}_{ix}}  = \ket{c_{ix} - u_x}$, $\ket{\bar{c}_{iy}}  = \ket{c_{iy} - u_y}$, and $\ket{\bar{c}_{iz}}  = \ket{c_{iz} - u_z}$.
	All the components of the vector $|{\bf c}_i|^0$ are equal to 1.
	The transformation matrix ${\bf M}$, whose components are constant, transforms the distribution functions into raw moments.
	The shift matrix {\bf N}, a lower-triangular matrix with components given by the macroscopic velocity ${\bf u}$, transforms the raw moments into CMs and can be written as ${\bf N} = {\bf T}{\bf M}^{-1}$.
	The practical forms of ${\bf M}$, ${\bf M}^{-1}$, ${\bf N}$, and ${\bf N}^{-1}$ are given in Appendix~\ref{sec:appA}.

The relaxation matrix ${\bf K}$ is a diagonal matrix given by
\begin{equation}
\begin{split}
		{\bf K} = {\rm diag} [
	&	s_0,s_1,s_1,s_1, s_{2\nu}, s_{2\nu},s_{2\nu}, s_{2\nu},s_{2\nu},s_{2b}, \\
	&	s_{3},s_{3},s_{3}, s_{3},s_{3},s_{3}, s_{3}, \\
	&	s_{4},s_{4},s_{4}, s_{4},s_{4},s_{4}, s_{5},s_{5},s_{5}, s_{6}],
\end{split}
\end{equation}
where the elements are the moments' relaxation times. If $s_0 = \ldots = s_i = \ldots = s_6$, the model reduces to the BGK (single-relaxation-time) model.
	The subscripts represent the orders (e.g., 2 means second-order).
	The parameters $s_{2\nu}$ and $s_{2b}$ satisfy the following relations:
\begin{align}
	\nu &= \frac{c^2}{3} \left(\frac{1}{s_{2\nu}} - \frac{1}{2} \right) \delta_t \label{eq:shear}
	\\
	\zeta &= \frac{5-3c^2}{9} \left(\frac{1}{s_{2b}} - \frac{1}{2} \right) \delta_t, \label{eq:bulk}
\end{align}
where $\nu$ and $\zeta$ are the kinematic and bulk viscosities, respectively.
	Here, we use $s_0 = s_1 = 0$ and $s_{2b}=s_3=s_4=s_5=s_6=1$.

	For the single-phase collision operator, we use the following enhanced equilibrium distribution function~\cite{Leclaire2013} in three-dimensions~\cite{Saito2017b}:
\begin{align}
	\notag
	f_i^{(e)}(\rho,{\bf u}) = & \rho \left \{ \varphi_i
	+ w_i 
	\left [\frac{3}{c^2} ({\bf c}_i \cdot {\bf u}) 
		+ \frac{9}{2c^4}({\bf c}_i \cdot {\bf u})^2 \right. \right. \\ 
		&- \left. \left. \frac{3}{2c^2}{\bf u}^2 
		+ \frac{9}{2c^6}({\bf c}_i \cdot {\bf u})^3
		- \frac{9}{2c^4}({\bf c}_i \cdot {\bf u}){\bf u}^2 \right] 
		 \right \} + \Phi_i.  \label{eq:EES}
\end{align}
	If $\Phi_i=0$, Eq.~(\ref{eq:EES}) reduces to the standard form of an equilibrium distribution function up to third order.
	Using Eq.~(\ref{eq:EES}) improves the Galilean invariance of the variable density and viscosity ratios under the assumption of a small pressure gradient~\citep{Leclaire2013,Leclaire2014,Leclaire2015}.

	The weights, $w_i$, are those of a standard D3Q27 lattice~\cite{He1997}, as follows:
\begin{equation}
	w_i = 
	\begin{cases}
		~8/27, & |{\bf c}_i|^2=0,\\
		~2/27, & |{\bf c}_i|^2=1, \\
		~1/54, & |{\bf c}_i|^2=2, \\
		~1/216, & |{\bf c}_i|^2=3.
\end{cases}
\end{equation}
	In addition, for a D3Q27 lattice, we can derive
\begin{equation}
	\varphi_i = 
	\begin{cases}
		~\bar{\alpha}, & |{\bf c}_i|^2=0,\\
		~2(1-\bar{\alpha})/19, & |{\bf c}_i|^2=1, \\
		~(1-\bar{\alpha})/38, & |{\bf c}_i|^2=2, \\
		~(1-\bar{\alpha})/152, & |{\bf c}_i|^2=3,
		\label{eq:varphi_ik}
\end{cases}
\end{equation}
and 
\begin{equation}
	\Phi_i = 
	\begin{cases}
		~-3 \bar{\nu}({\bf u}\cdot \nabla \rho)/c, & |{\bf c}_i|^2=0,\\
		~+16\bar{\nu}({\bf G} : {\bf c}_i \otimes {\bf c}_i)/c^3, & |{\bf c}_i|^2=1, \\
		~+4 \bar{\nu}({\bf G} : {\bf c}_i \otimes {\bf c}_i)/c^3, & |{\bf c}_i|^2=2, \\
		~+1 \bar{\nu}({\bf G} : {\bf c}_i \otimes {\bf c}_i)/c^3, & |{\bf c}_i|^2=3, \label{eq:Phi}
\end{cases}
\end{equation}
where $\otimes$ is the tensor product, ``$:$'' represents tensor contraction, and 
$\bar{\nu}$ is the kinematic viscosity, which interpolates between the red and blue viscosities $\nu_r$ and $\nu_b$ via the following harmonic mean~\cite{Liu2014,Zu2013,Leclaire2017}:
\begin{equation}
	\frac{1}{\bar{\nu}} = \frac{1+\phi}{2}\frac{1}{\nu_r} + \frac{1-\phi}{2}\frac{1}{\nu_b}.
\end{equation}
Here, $\phi$ is the order parameter that distinguishes the two components in the multicomponent flow, defined as~\cite{Ba2016}
\begin{equation}
	\phi = \left(\frac{\rho_r}{\rho_r^0}-\frac{\rho_b}{\rho_b^0} \right) \bigg/ \left(\frac{\rho_r}{\rho_r^0}+\frac{\rho_b}{\rho_b^0} \right),
	\label{eq:order}
\end{equation}
	where the superscript ``0'' indicates the initial density value at the beginning of the simulation~\citep{Leclaire2013}.
	The order parameter value $\phi=1,-1$, and $0$ correspond to a purely red fluid, a purely blue fluid, and the interface between the two, respectively~\cite{Tolke2002}.
	In the D3Q27 lattice framework, the tensor ${\bf G}$ in Eq.~(\ref{eq:Phi}) is defined as
\begin{equation}
	{\bf G} = \frac{1}{48} \left[{\bf u}\otimes \nabla \rho
	+ ({\bf u}\otimes \nabla \rho)^{\top} \right].
\end{equation}
	As established in Ref.~\cite{Grunau1993}, in order to obtain a stable interface, we must take the fluid density ratio $\gamma$ into account, which is defined as follows:
\begin{equation}
	\gamma = \frac{\rho_r^0}{\rho_b^0} = \frac{1-\alpha_b}{1-\alpha_r}.
\end{equation}
	The fluid pressures are given by an isothermal equation of state for the D3Q27 lattice:
\begin{equation}
p = \rho \left(c_s \right)^2 = \rho_k \frac{9(1-\bar{\alpha})}{19}c^2,
	\label{eq:press}
\end{equation}
where $\bar{\alpha}$ interpolates between $\alpha_r$ and $\alpha_b$ as follows~\cite{Leclaire2017}
\begin{equation}
	\bar{\alpha} = \frac{1 + \phi}{2} \alpha_r + \frac{1 - \phi}{2} \alpha_b.
\end{equation}
	In this paper, we set $\alpha_b = 8/27$, for which $c_s^b = 1/\sqrt{3}$~\cite{Leclaire2014,Saito2016}.

	The term $\Ket{F}$ in Eq.~(\ref{eq:scoll}) is a discrete forcing term that accounts for the body force ${\bf F}$. 
	In the GMRT framework~\citep{Fei2017}, it is
\begin{equation}
	\Ket{F} = {\bf M}^{-1} {\bf N}^{-1} \left({\bf I} - \frac{1}{2} {\bf K} \right) {\bf N} {\bf M} \Ket{F'},
	\label{eq:ketF}
\end{equation}
where ${\bf I}$ is the unit matrix, $\ket{F} = (F_0,F_1,\dots,F_{26})^{{\top}}$, and $\ket{F'} = (F'_0,F'_1,\dots,F'_{26})^{{\top}}$ is given by
\begin{equation}
	\Ket{F'} = w_i \left[3\frac{{\bf c}_i - {\bf u}}{c^2} 
	+ 9\frac{({\bf c}_i \cdot {\bf u}) {\bf c}_i}{c^4} \right] \cdot {\bf F}\delta_t.
	\label{eq:ketFd}
\end{equation}
	Equations~(\ref{eq:ketF}) and (\ref{eq:ketFd}) reduce to the MRT forcing scheme~\cite{Yu2010} when we use ${\bf N} = {\bf I}$,
	and to Guo {\it et al}.'s original forcing scheme~\cite{Guo2002} when we use a single-relaxation time. 
	It is not necessarily required that ${\bf N} = {\bf M} = {\bf I}$ to degrade into Guo {\it et al}.'s scheme.

	To model the interfacial tension, we use the generalized perturbation operator derived in Ref.~\citep{Liu2012}, based on the idea of continuum surface force (CSF)~\cite{Brackbill1992}, and follow~\citep{Reis2007} to obtain the interfacial tension as follows:
\begin{equation}
	\left(\Omega_i \right)^{(2)} = \frac{A}{2} |\nabla \phi| 
	\left[w_i \frac{({\bf c}_i \cdot \nabla \phi)}{|\nabla \phi|^2} - B_i \right].
	\label{eq:perturb}
\end{equation}
	Equation~(\ref{eq:perturb}) takes the correct form for an interfacial tension force in the Navier--Stokes equations when the lattice-specific variables $B_i$ are chosen correctly.
	We have derived the following $B_i$ values for the D3Q27 lattice framework:
\begin{equation}
	B_i = 
	\begin{cases}
		~-(10/27)c^2, & |{\bf c}_i|^2=0,\\
		~+(2/27)c^2, & |{\bf c}_i|^2=1, \\
		~+(1/54)c^2, & |{\bf c}_i|^2=2, \\
		~+(1/216)c^2, & |{\bf c}_i|^2=3. \label{eq:B_i}
\end{cases}
\end{equation} 
	In this model, the interfacial tension can be given directly by
\begin{equation}
	\sigma = \frac{4}{9} A \tau c^4 \delta_t, \label{eq:sigma}
\end{equation}
where $\tau$ is the relaxation time and we have assumed that $A=A_r=A_b$. 
The parameter $A$ controls the interfacial tension strength $\sigma$.

	Although the perturbation operator $(\Omega_i^k)^{(2)}$ generates the interfacial tension, it does not guarantee the two fluids are immiscible. 
	To promote phase segregation and maintain the interface, we apply the following recoloring operators~\cite{Latva-Kokko2005,Halliday2007,Leclaire2012}
\begin{eqnarray}
	(\Omega_i^r)^{(3)} = \frac{\rho_r}{\rho}f_i 
	+ \beta \frac{\rho_r \rho_b}{\rho^2} \cos(\theta_i)
	f_i^{(e)}(\rho,\bf{0}), \label{eq:op3r}
\\
	(\Omega_i^b)^{(3)} = \frac{\rho_b}{\rho}f_i 
	- \beta \frac{\rho_r \rho_b}{\rho^2} \cos(\theta_i)
	f_i^{(e)}(\rho,\bf{0}), \label{eq:op3b}
\end{eqnarray}
where $\theta_i$ is the is the angle between $\nabla\phi$ and ${\bf c}_i$, defined by
\begin{equation}
	\cos(\theta_i) = \frac{{\bf c}_i \cdot \nabla \phi}{|{\bf c}_i|  |\nabla \phi|}.
\end{equation}
	Here, we set the parameter $\beta$ to $0.7$ to reproduce the correct interfacial behavior with as narrow an interface as possible~\citep{Halliday2007,Liu2012,Liu2017}.

	For the current model, we can derive the following continuity and Navier--Stokes equations via Chapman--Enskog analysis~\cite{Liu2012,Guo2002,Kruger2017}
\begin{align}
	\frac{\partial \rho}{\partial t} &+ \nabla \cdot (\rho {\bf u}) = 0, \label{Continuity}
	\\
	\frac{\partial (\rho {\bf u})}{\partial t} + \nabla \cdot (\rho {\bf u} {\bf u})
	&= -\nabla p + \nabla \cdot {\bf \Pi} + \nabla \cdot {\bf S} + {\bf F}, \label{N-S}
\end{align}
where 
\begin{equation}
	{\bf \Pi} = \rho \nu [\nabla {\bf u} + (\nabla {\bf u})^{{\rm T}}] + \rho (\zeta - 2\nu/D)(\nabla \cdot {\bf u}){\bf I}
\end{equation}
is the viscous stress tensor, with $D=3$ in the three dimensions; the shear viscosity $\nu$ is given by Eq.~(\ref{eq:shear}) and the bulk viscosity $\zeta$ is given by Eq.~(\ref{eq:bulk}).
	In Eq.~(\ref{N-S}), the $\displaystyle \nabla \cdot {\bf S}$ term arises from the perturbation operator given by Eq.~(\ref{eq:perturb}) and, according to the CSF idea, is equivalent to the interfacial force~\cite{Liu2012}.
	The capillary stress tensor ${\bf S}$ is given by
\begin{equation}
	{\bf S} = -\tau \delta_t \sum_i \sum_k \left(\Omega_i^k \right)^{(2)} {\bf c}_i {\bf c}_i.
\end{equation}

	The solutions of the present model with CMs~\citep{DeRosis2017a} are consistent with the Navier--Stokes equations to second order in diffusive scaling~\citep{Liu2016,Geier2015,Lycett-Brown2014,DeRosis2016} with the body~\cite{Guo2002} and interfacial~\citep{Liu2012} forces.

	To compute the gradient operator for an arbitrary function $\chi$, 
we adopt the following second-order isotropic central scheme~\cite{Liu2012,Guo2011,Liang2014,Lou2012}:
\begin{equation}
	\nabla \chi({\bf x},t) = \frac{3}{c^2} \sum_i \frac{w_i \chi({\bf x}+{\bf c}_i \delta_t,t) {\bf c}_i}{\delta_t}.
\end{equation}
	In this paper, we set $\delta_x$ and $\delta_t$ to 1, as is usual in LB simulations.
	Although the above formulation focuses on two-component systems,
	it should also be straightforward to implement this model for systems with three or more components~\citep{Leclaire2013b}.

\section{Static droplet tests \label{sec:laplace}}

	In this section, we carry out static droplet tests to evaluate whether the interfacial tension predicted by Eq.~(\ref{eq:sigma}) is correct for various density ratios.
	We discretized the computational domain as a $100\times100\times100$ lattice and immersed a static red droplet of radius $R$ in a blue fluid.
	The initial density fields for each phase were as follows:
\begin{align}
	\rho_r(x,y,z) = \frac{\rho_r^0}{2} \left[1 -  \tanh 
	\left( \frac{2(r-R)}{W} \right) \right],
\\
	\rho_b(x,y,z) = \frac{\rho_b^0}{2} \left[1 +  \tanh 
	\left( \frac{2(r-R)}{W} \right) \right],
\end{align}
where $W=4$ and $r=\sqrt{(x-x_c)^2+(y-y_c)^2+(z-z_c)^2}$. Here $(x_c,y_c,z_c)$ is the center of the computational domain.
	We set the kinematic viscosity ratio to be 1, with each phase's kinematic viscosity being $1/6$, and set the parameters $A$ in Eq.~(\ref{eq:sigma}) to 0.01 and the initial droplet radius $R$ to 25.
	We neglect gravity throughout the simulations and imposed periodic boundary conditions on all sides of the computational domain.

The three-dimensional Laplace equation is given by 
\begin{equation}
	\Delta p = \frac{2\sigma}{R},
	\label{eq:lap}
\end{equation}
where $\Delta p$ is the pressure difference across the droplet interface.
	We evaluated the pressure for each phase using Eq.~(\ref{eq:press}) and measured it after $100\,000$ iterations via the procedure used by \citet{Leclaire2012}.
	Table~\ref{tab:laplace} summarizes the simulation parameters and resulting errors $E$, which were calculated as~\citep{Liu2012}
\begin{equation}
	E = \frac{|\sigma_{\mathrm{th}} - \sigma_{\mathrm{Lap}}|}{\sigma_{\mathrm{th}}} \times 100,
\end{equation}
	where $\sigma_{\mathrm{th}}$ and $\sigma_{\mathrm{Lap}}$ are the interfacial tensions predicted by Eq.~(\ref{eq:sigma}) and measured using the Laplace equation [Eq.~(\ref{eq:lap})], respectively.
	Note that the droplets are always spherical at equilibrium, indicating that numerical stability was maintained for all density ratios.
	These tests confirm that the CM-based CG model described in Sec.~\ref{sec:methodology} can predict the interfacial tension in static cases to within a maximum error of 0.40\%.
	In addition, we measured the maximum spurious velocity $|{\bf u}_{\mathrm{max}}|$ in the domain at equilibrium, finding a maximum value of $1.22\times10^{-4}$ (Table~\ref{tab:laplace}).
	This value is smaller than that of our MRT-based CG model~\citep{Saito2017b} ($|{\bf u}_{\mathrm{max}}|=5.8\times 10^{-3}$ for $\gamma=1.5$).
	These findings indicate that introducing the CMs into the single-phase collision operator can help to reduce the spurious velocity, contributing to enhance the numerical stability.

\begingroup
\renewcommand{\arraystretch}{1.4}
\begin{table*}[tb]
\caption{\label{tab:laplace} Static droplet tests for various density ratios, 
	showing that the current CM-based CG model can predict the interfacial tension for static cases with a maximum error of 0.40\% and
	that introducing CMs into the single-phase collision operator can help to reduce the spurious velocity.}
\begin{ruledtabular}
\begin{tabular}{cccccc}
Density ratio $\gamma$& Theoretical ($\sigma_{\mathrm{th}}$) & Numerical ($\sigma_{\mathrm{Lap}}$) & Error $E$ (\%) & Maximum spurious velocity $|{\bf u}_{\mathrm{max}}|$  \\
\colrule
$1$    & $3.5556 \times 10^{-4}$ & $3.5697\times 10^{-4}$ & 0.40 & $1.22\times 10^{-4}$ \\
$10$   & $3.5556 \times 10^{-4}$ & $3.5683\times 10^{-4}$ & 0.36 & $4.23\times 10^{-5}$ \\
$100$  & $3.5556 \times 10^{-4}$ & $3.5669\times 10^{-4}$ & 0.32 & $4.67\times 10^{-5}$ \\
$1000$ & $3.5556 \times 10^{-4}$ & $3.5696\times 10^{-4}$ & 0.40 & $6.94\times 10^{-5}$ 
\end{tabular}
\end{ruledtabular}
\end{table*}
\endgroup

\section{Jet breakup simulations \label{sec:jetBreakup}}

	As mentioned in Sec.~\ref{sec:intro}, we will now simulate the following two experiments using the method presented in Sec.~\ref{sec:methodology}:
	\begin{itemize}
	\item UT (University of Tsukuba) experiments~\citep{Matsuo2013} (see Sec.~\ref{sec:UT})
	\item FT (FARO-TERMOS) experiments~\citep{Magallon1992}  (see Sec.~\ref{sec:FT})
	\end{itemize}
	
	In the following simulations, we neglect temperature changes and do not take phase-change effects (e.g., vaporization, condensation, or solidification) into account, meaning that these are strictly hydrodynamic simulations.
	
\subsection{Setup}
\begin{figure}[tb]
	\includegraphics[width=8.5cm]{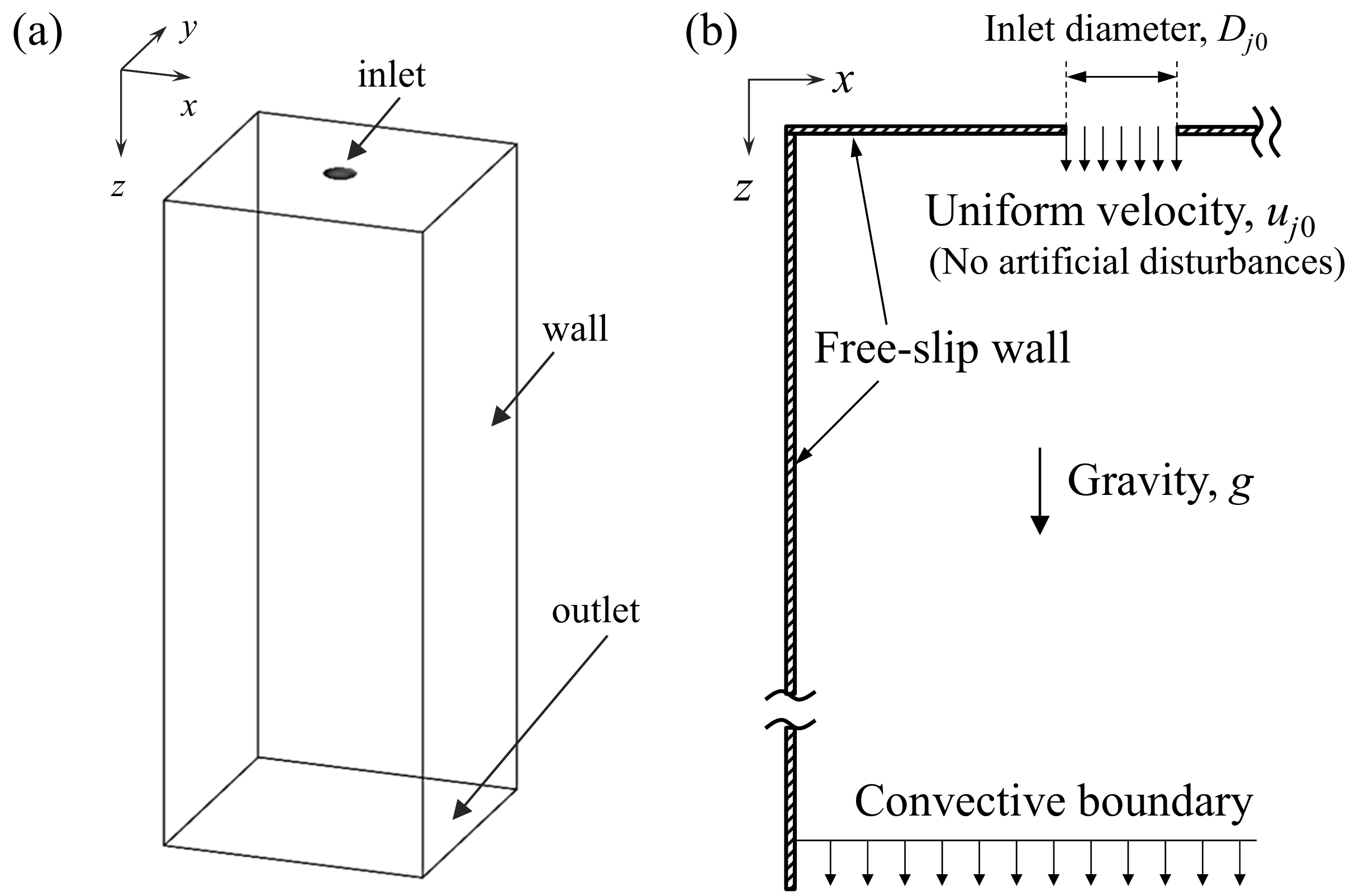}
	\caption{Boundary conditions for the melt-jet breakup simulations. 
	(a) The boundaries consist of an inflow boundary, wall boundaries, and an outflow boundary. 
	(b) There is a circular inflow boundary of diameter $D_{j0}$ at the top, where the velocity $u_{j0}$ is uniform.
	Free-slip~\citep{Succi2001} and convective~\citep{Lou2013} boundary conditions are imposed at the wall and outflow boundaries, respectively.
	\label{fig:boundaryJet}}
\end{figure}

	Figure~\ref{fig:boundaryJet} illustrates the computational setup for our hydrodynamic melt-jet breakup simulations.
   The boundary conditions are the same as in Ref.~\citep{Saito2017b}. 
	Initially, the computational domain consists entirely of blue particle distribution functions $f_i^b$ with zero velocity.
	The boundaries consist of an inflow boundary, wall boundaries, and an outflow boundary.
	There is a circular inflow boundary at the top of the domain, within $(x-x_c)^2 + (y-y_c)^2 < (D_{j0}/2)^2$, where $(x_c,y_c)$ represents the center of the $x$-$y$ plane.
	Here, the velocity $u_{j0}$ is uniform, with corresponding equilibrium functions, and there are no artificial disturbances at this boundary.
	Wall boundaries cover the rest of the top and sides of the domain, with free-slip~\citep{Succi2001} boundary conditions.
	At the outflow boundary, we imposed a convective boundary condition~\citep{Lou2013}, applying the following convective equation to the distribution functions:
\begin{equation}
	\frac{\partial f_i}{\partial t} + U_c \frac{\partial f_i}{\partial z} = 0, ~ {\rm at} ~z=N,
\end{equation} 
where $N$ is the outflow boundary node.
	Following \citet{Lou2013}, we added two additional ghost nodes, $N+1$ and $N+2$.
	The discretized form of the distribution functions can be given by the first-order implicit scheme
\begin{align}
	f_i(x,y,N,t+\delta_t) 
	= \frac{f_i(x,y,N,t)+\lambda f_i(x,y,N-1,t+\delta_t)}{1+\lambda},
\end{align}
where $\lambda=U_c(t+\delta_t)\delta_t/\delta_x$.
	There are several possibilities for the convective velocity $U_c$ normal to the outflow boundary, such as the local, average, or maximum velocity~\citep{Orlanski1976}.
	After conducting some numerical tests, we determined that the local velocity was most suitable for the current system, namely,
\begin{equation}
	U_c(x,y,N,t) = u_z (x,y,N-1,t),
\end{equation}
where $u_z({\bf x},t) = u_z (x,y,z,t)$ is the component of the fluid velocity ${\bf u}$ in the $z$-direction.

	We represented the body force in Eq.~(\ref{eq:ketFd}) as 
\begin{equation}
	{\bf F}({\bf x},t) = \left(\rho({\bf x},t)-\rho_b^0 \right){\bf g}, 
	\label{eq:bodyForceJet}
\end{equation}
with ${\bf g} = (0,0,g)$.
	This means that gravity only acts on the dispersed phase~\citep{Chen2014}.

\subsection{UT experiments \label{sec:UT}}

	In the UT experiments~\cite{Matsuo2013}, an alloy called U-Alloy78 (Osaka Asahi Co., Ltd.), with a melting point of 78 $\rm ^oC$, was injected into a stagnant water pool under atmospheric pressure.
	We considered four cases with different nozzle diameters ($D_{j0}$ = 7, 10, 15, and 20 mm) from Ref.~\citep{Matsuo2013} (see Table~\ref{tab:UT}).
	In these four cases, the melt and water temperatures were set to 270$\mathrm{^oC}$ and 70$\mathrm{^oC}$, respectively,
	and the physical properties were as follows: $\rho_j=8\,183~\mathrm{kg/m^3}$, $\nu_j=0.24~\mathrm{mm^2/s}$, $\sigma=1.104~\mathrm{N/m}$, $\rho_c=981~\mathrm{kg/m^3}$, and $\nu_c=0.443~\mathrm{mm^2/s}$.
	Note that they defined the jet velocity $u_{j0}$ and inlet diameter $D_{j0}$ as the contact velocity at the water surface and the nozzle diameter, respectively.
	
\begingroup
\renewcommand{\arraystretch}{1.4}
\begin{table*}[tb]
\caption{\label{tab:UT} 
Conditions used for the UT experiment simulations, reproduced from Ref.~\citep{Matsuo2013}. 
The dimensionless parameters calculated from Eqs.~(\ref{eq:densityRatio})--(\ref{eq:Froude}) are also shown,
including the density ratio $\gamma$, kinematic viscosity ratio $\eta$, Reynolds number $\mathrm{Re}$, Weber number $\mathrm{We}$, and Froude number $\mathrm{Fr}$.}
\begin{ruledtabular}
\begin{tabular}{cccccccc}
 & 
 $D_{j0}~[\mathrm{mm}]$ & 
 $u_{j0}~[\mathrm{m/s}]$ & 
 $\gamma~[-]$ & 
 $\eta~[-]$ & 
 $\mathrm{Re}~[-]$ & 
 $\mathrm{We}~[-]$ & 
 $\mathrm{Fr}~[-]$ \\ 
\colrule
Case 1 & $7 $ & $2.10$ & $8.3$ & $0.54$ & $6.1\times10^4$ & $2.3\times10^2$ & 64\\
Case 2 & $10$ & $1.55$ & $8.3$ & $0.54$ & $6.5\times10^4$ & $1.8\times10^2$ & 24\\
Case 3 & $15$ & $1.73$ & $8.3$ & $0.54$ & $1.1\times10^5$ & $3.3\times10^2$ & 20\\
Case 4 & $20$ & $1.75$ & $8.3$ & $0.54$ & $1.5\times10^5$ & $4.5\times10^2$ & 16\\
\end{tabular}
\end{ruledtabular}
\end{table*}
\endgroup
	
	In this case, we discretized the computational domain into an $8D_{j0}\times8D_{j0}\times40D_{j0}$ lattice with $D_{j0} = 30$,
	resulting in a total of $240\times240\times1\,200=69\,120\,000$ grid points being used in the simulation.
	We set the inlet velocity $u_{j0}$ to 0.05, the jet density $\rho_{j}=\rho_r^0$ to 1, and the coolant density $\rho_{c}=\rho_b^0$ to $1/\gamma$.
	We here mention the computational cost of the present simulations. 
	For the simulations with 6\,912\,000 grids, it takes around 32 hours for 20\,000 iterations with our computational environment. One way to reduce the cost may be using the reduced velocity model, such as D3Q15 or D3Q19 lattice models (see Refs.~\citep{Fei2018b,DeRosis2017a}).
	
\begin{figure}[tb]
\begin{center}
	\includegraphics[width=6.4cm]{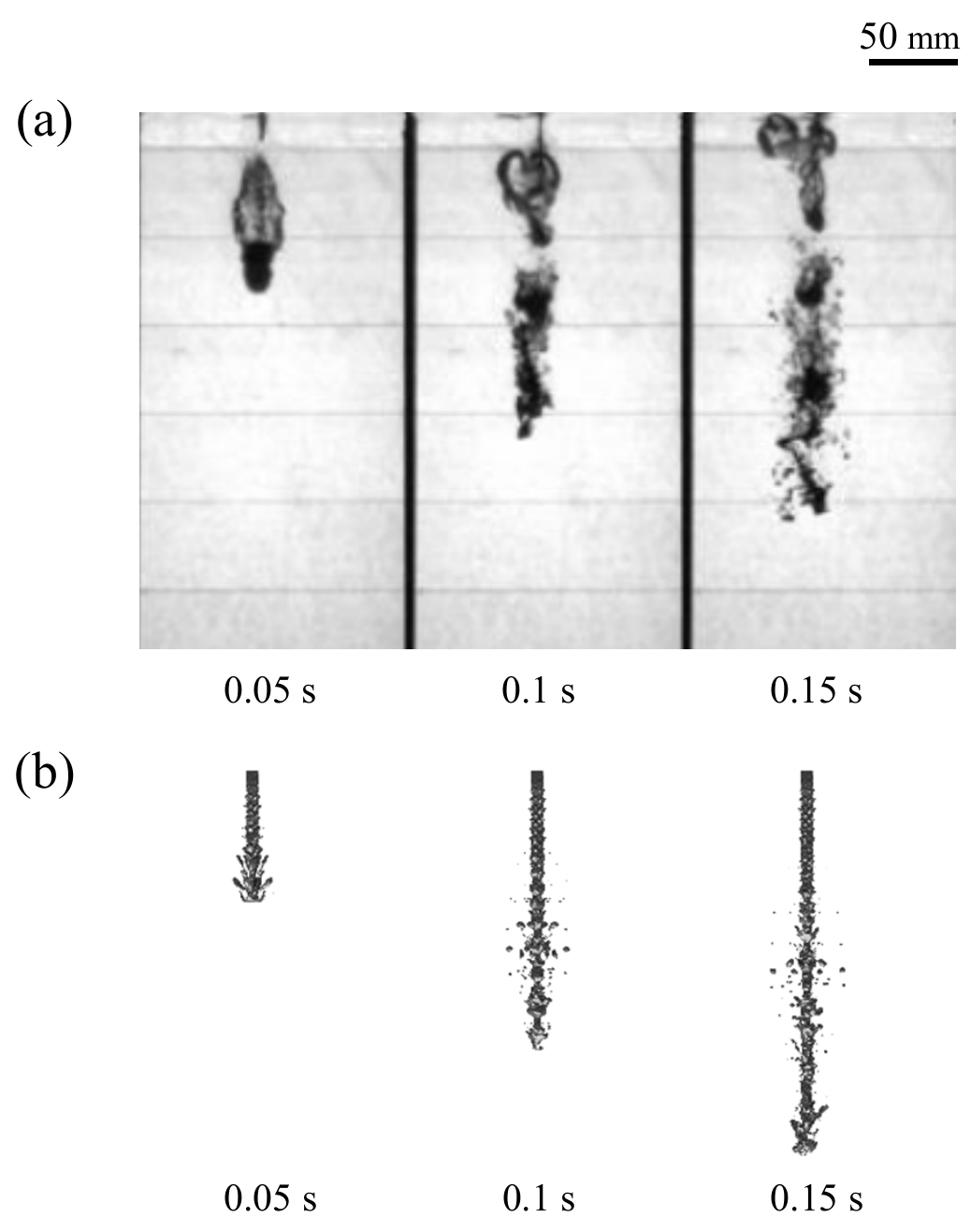}
	\caption{Comparison of jet breakup behavior for Case 1 in Table~\ref{tab:UT} ($D_{j0}=7$ mm), for the (a) UT experiment and (b) simulation. 
	The simulation parameters used were as follows (in lattice units): $\sigma=6.6\times10^{-3}$, $\nu_j=\nu_r=2.5\times10^{-5}$, $\nu_c=\nu_b=4.5\times10^{-5}$, $g=1.3\times10^{-6}$, $\rho_j=\rho_r^0=1$, and $\rho_c=\rho_b^0=0.12$.
	The minimum spatial resolution in this case was $\Delta x= 0.23$ mm.
	The experimental results clearly show trapped gas, which caused differences in the interface shape compared with the simulation.
	\label{fig:jet_d07}}
\end{center}
\end{figure}

\begin{figure}[tb]
\begin{center}
	\includegraphics[width=7.0cm]{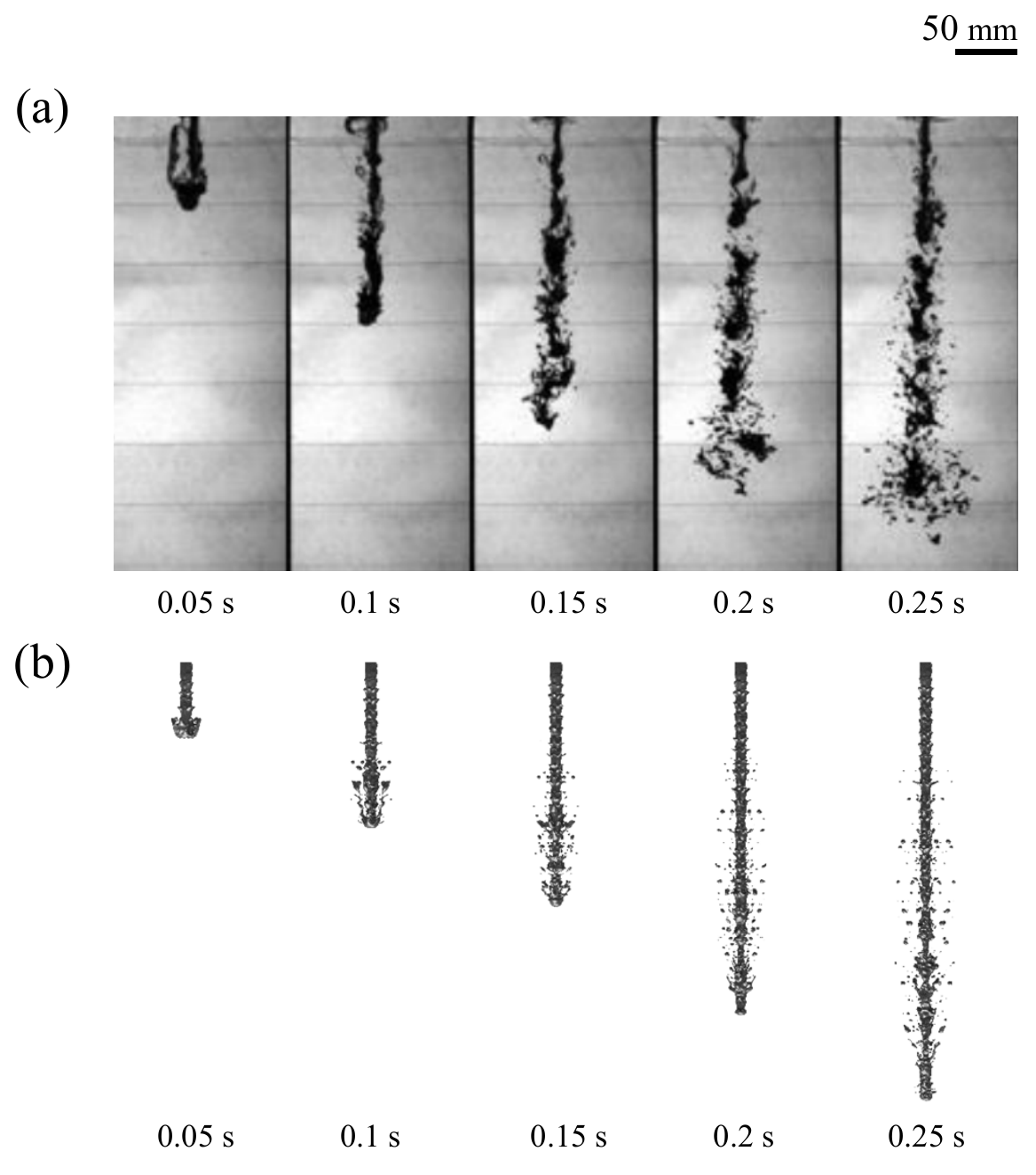}
	\caption{Comparison of jet breakup behavior for Case 2 in Table~\ref{tab:UT} ($D_{j0}=10$ mm), for the (a) UT experiment and (b) simulation. 
	The simulation parameters used were as follows (in lattice units): $\sigma=8.4\times10^{-3}$, $\nu_j=\nu_r=2.3\times10^{-5}$, $\nu_c=\nu_b=4.3\times10^{-5}$, $g=3.4\times10^{-6}$, $\rho_j=\rho_r^0=1$, and $\rho_c=\rho_b^0=0.12$.
	The minimum spatial resolution in this case was $\Delta x = 0.33$ mm.
	Between $t=0.2$ and $t=0.25$ s, the fragments in the simulation spread out significantly less than in the experiments. 
	\label{fig:jet_d10}}
\end{center}
\end{figure}

\begin{figure}[tb]
\begin{center}
	\includegraphics[width=6.5cm]{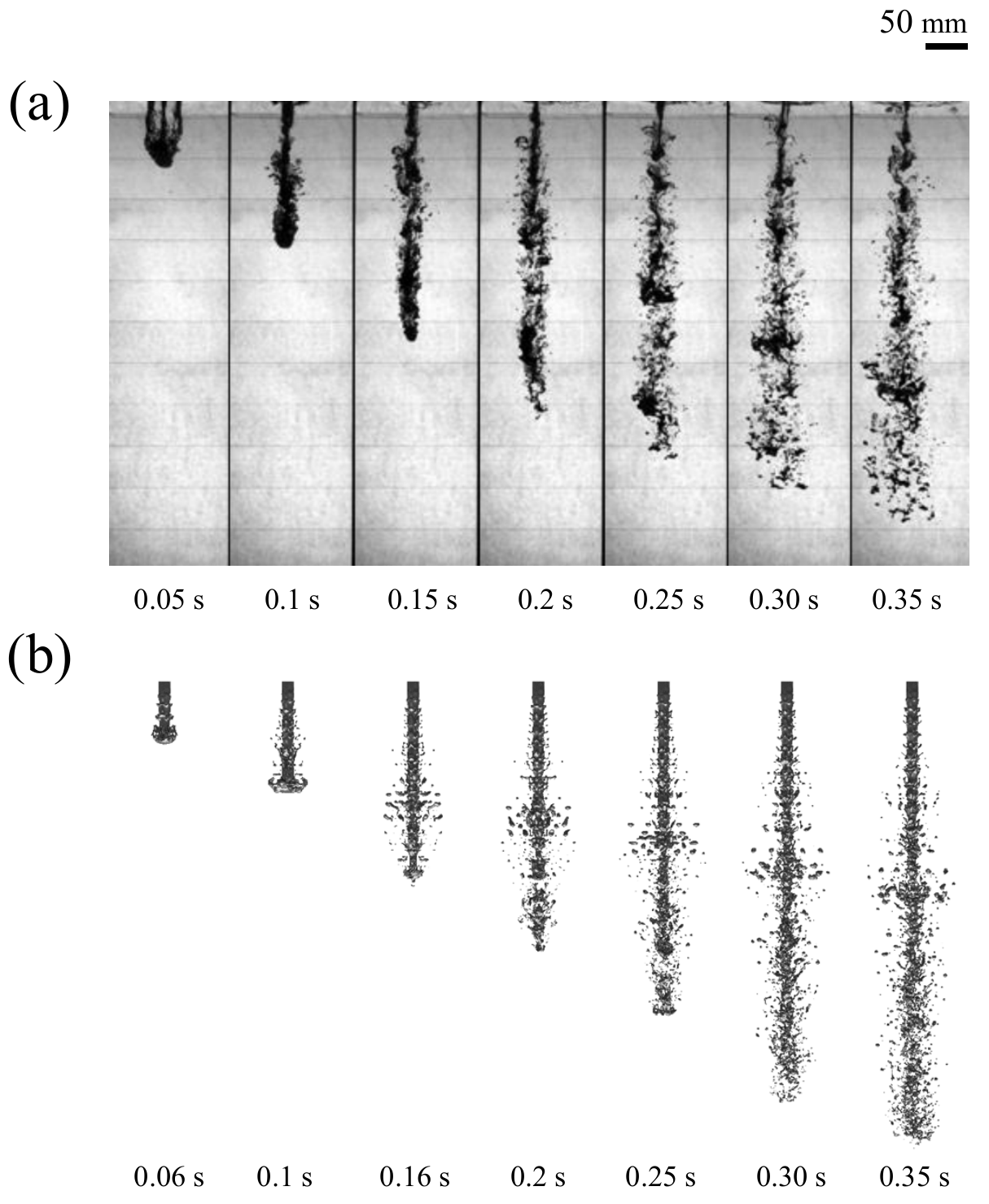}
	\caption{Comparison of jet breakup behavior for Case 3 in Table~\ref{tab:UT} ($D_{j0}=15$ mm), for the (a) UT experiment and (b) simulation. 
	The simulation parameters used were as follows (in lattice units): $\sigma=4.5\times10^{-3}$, $\nu_j=\nu_r=1.4\times10^{-5}$, $\nu_c=\nu_b=2.6\times10^{-5}$, $g=4.1\times10^{-6}$, $\rho_j=\rho_r^0=1$, and $\rho_c=\rho_b^0=0.12$.
	The minimum spatial resolution in this case was $\Delta x = 0.5$ mm.
	Although the results spread out more in the lateral direction than the experimental results do between $t=0.16$ and $0.25$ s, their final shape is close to that observed experimentally.
	\label{fig:jet_d15}}
\end{center}
\end{figure}

\begin{figure}[tb]
\begin{center}
	\includegraphics[width=8.0cm]{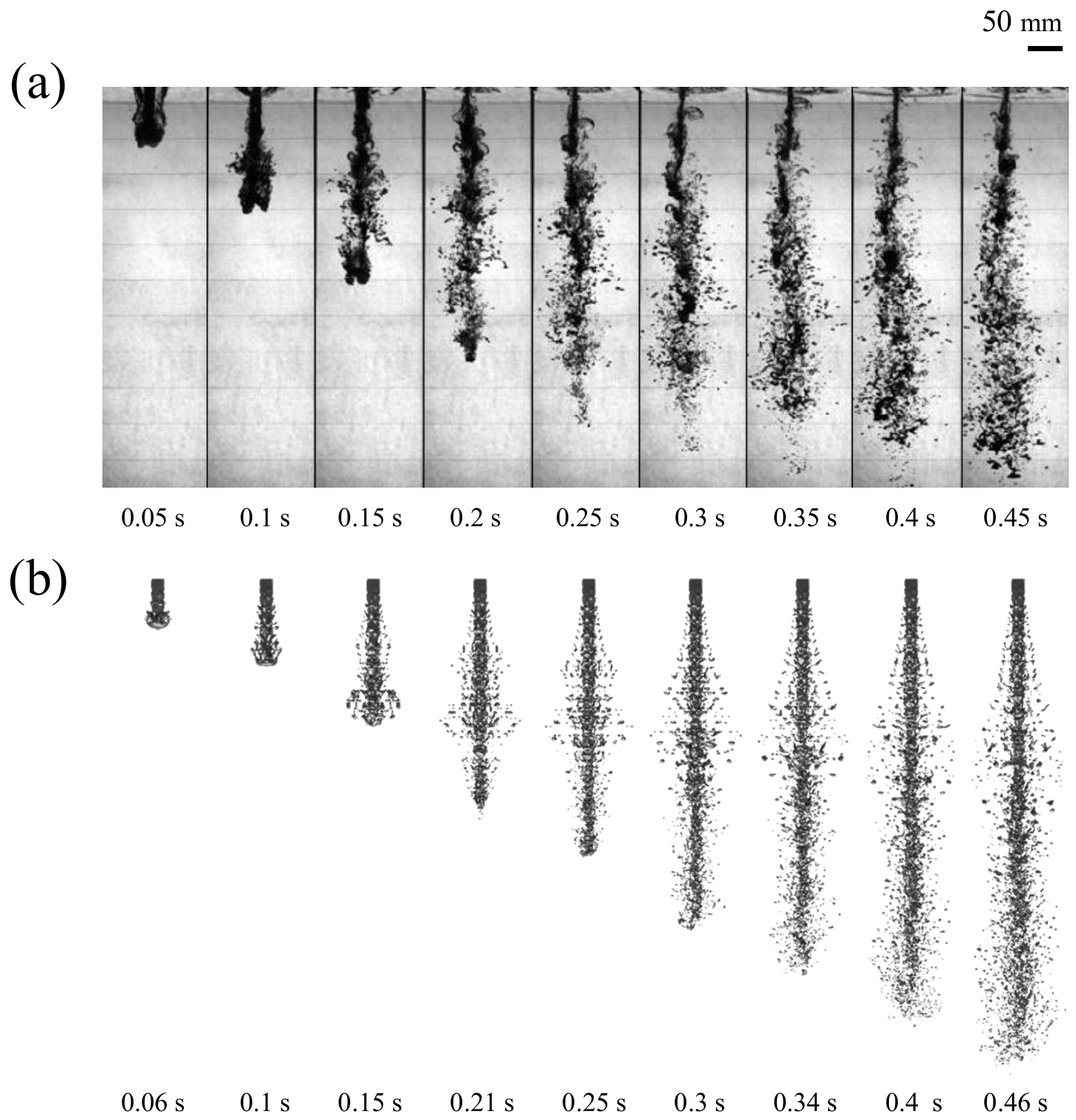}
	\caption{Comparison of jet breakup behavior for Case 4 in Table~\ref{tab:UT} ($D_{j0}=20$ mm), for the (a) UT experiment and (b) simulation. 
	The simulation parameters used were as follows (in lattice units): $\sigma=3.3\times10^{-3}$, $\nu_j=\nu_r=1.0\times10^{-5}$, $\nu_c=\nu_b=1.9\times10^{-5}$, $g=5.3\times10^{-6}$, $\rho_j=\rho_r^0=1$, and $\rho_c=\rho_b^0=0.12$.
	The minimum spatial resolution in this case was $\Delta x = 0.67$ mm.
	In the simulation, the fragment groups were generated around the jet's leading edge and
	the side of the jet column, showing similar behavior to that observed in the experiment.
	\label{fig:jet_d20}}
\end{center}
\end{figure}

\begin{figure*}[tb]
\begin{center}
	\includegraphics[width=16cm]{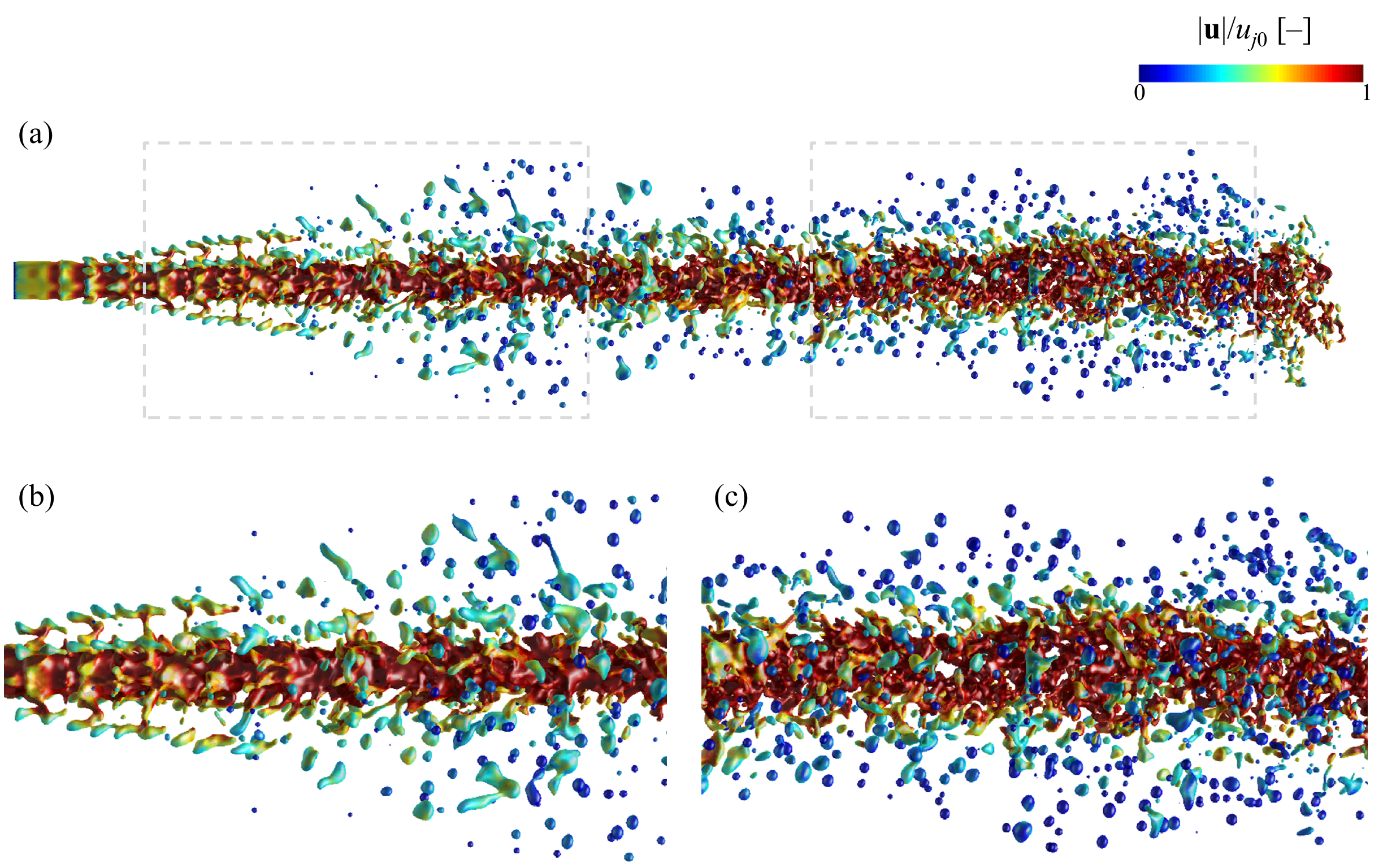}
	\caption{(a) Snapshot of detailed interface structure for Case 4 in Table~\ref{tab:UT} ($D_{j0}=20$ mm).
	The color bar indicates the velocity magnitude normalized by the inlet velocity.
	The liquid column has large velocity, while the generated fragments has small velocity.
	The regions surrounded by gray dashed lines are shown below.
	(b) Magnified snapshot for upstream region.
	The fragments generate from the unstable liquid jet interface.
	Most of the fragments in this region are stretched, which appear not to be spherical shapes.
	The velocity magnitude of stretched fragments is large, while that of spherical ones is small.
	(c) Magnified snapshot for downstream region.
	Most of the fragments in this region are spherical shapes with low velocity magnitude.
	\label{fig:interfaceVelocity}}
\end{center}
\end{figure*}

	Figures~\ref{fig:jet_d07}--\ref{fig:jet_d20} show comparisons of the UT experiments' results with those of our simulations for Cases 1--4.
	The simulation parameters used are shown (in lattice units) in the captions.
	All the simulations (Table~\ref{tab:UT}) were numerically stable, even for very low kinematic viscosities of $O(10^{-5})$.
	In all cases, the simulation results reproduce the qualitative interfacial behavior well, i.e., many fragments are generated as the jets penetrate each other, both around the leading edge and the side regions.
	However, there are still some differences in interfacial behavior between the experiments and the simulations, particularly for smaller $D_{j0}$ values (e.g., at $t=0.15$ s in Fig.~\ref{fig:jet_d07} and $t=0.25$ s in Fig.~\ref{fig:jet_d10}).
	One reason for this is the effect of gas entrapment:
	in the experiments, some of the gas was trapped in the water pool when the jet came in contact with the water surface,
	and the simulations did not take this into account.
	In contrast, the simulated shape of the interface agreed relatively well with the experiments for larger $D_{j0}$ values (Figs.~\ref{fig:jet_d15} and \ref{fig:jet_d20})  because little gas was trapped in these cases.
	For Case 4, the details of the generated fragments and the flow field are shown in Fig.~\ref{fig:interfaceVelocity}.
	We can find that the liquid jet column has large velocity, while the generated fragments has small velocity.
	In the snapshot at upstream region [Fig.~\ref{fig:interfaceVelocity}(b)], the fragments generate from the unstable liquid-jet interface.
	Most of the fragments in this region are stretched, which appear not to be spherical shapes.
	The velocity magnitude of stretched fragments are large, while that of spherical ones are small.
	In the snapshot at downstream region [Fig.~\ref{fig:interfaceVelocity}(c)], most of the fragments are spherical shapes with low velocity magnitude.
	
\begin{figure}[tb]
\begin{center}
	\includegraphics[width=9cm]{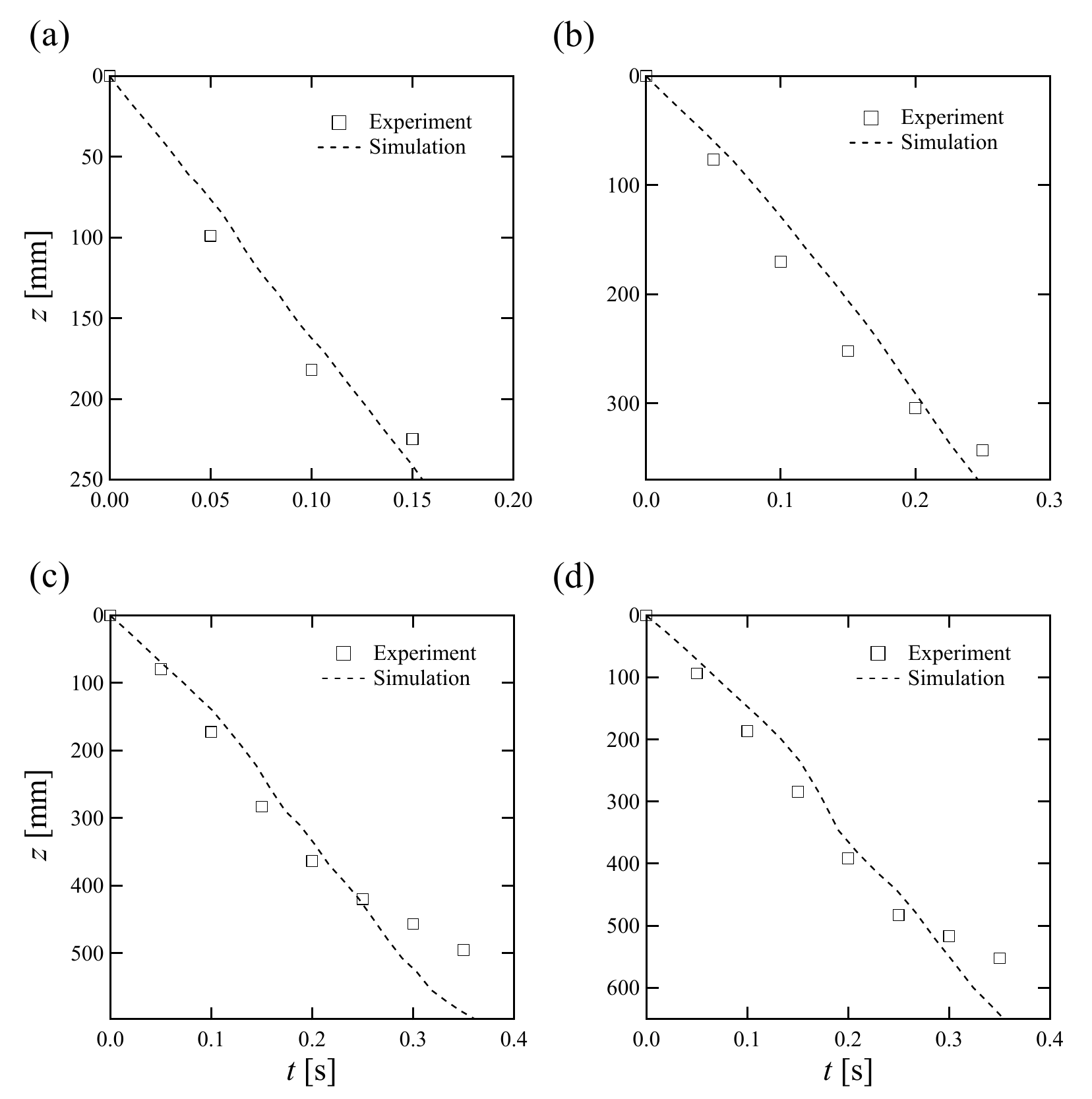}
	\caption{Evolution of the jet's leading edge over time for (a) Case 1 ($D_{j0}=7$ mm), (b) Case 2 ($D_{j0}=10$ mm), (c) Case 3 ($D_{j0}=15$ mm), and (d) Case 4 ($D_{j0}=20$ mm).
	In all cases, the experimental observations are ahead of the simulation early on, but the simulation passes the experiment in the latter stages.
	Overall, the simulations reproduce the experimental trends well.
	\label{fig:leadEdge}}
\end{center}
\end{figure}

	In order to make a quantitative comparison, we compared the evolution of the jet's leading edge over time for each condition, and 
	the results are summarized in Fig.~\ref{fig:leadEdge}.
	This shows that the simulation results agree well with the experimental data~\citep{Matsuo2013}. 
	In all cases, the experimental observations are ahead of the simulation early on, 
	but the simulation passes the experiment in the later stages. 
	This is due to differences in the inlet conditions between the experiments and simulations: 
	in the experiments, a given mass of melt material is injected, so the injection velocity $u_{j0}$ may decrease over time (i.e., is not constant), whereas the simulations assumed a constant injection velocity. 
	This means that the simulations match the experiments particularly well in the early stages, during jet penetration.
	
	Based on the evolution of the jet's leading edge over time, we can estimate the jet breakup length $L$, i.e., the length of the continuous liquid column emitted from the nozzle~\citep{McCarthy1974,Lin1998}.
	This is one of the metrics that characterize jet breakup behavior.
	Here, we estimated it via the procedure used by previous melt-jet experiments~\citep{Magallon1997,Matsuo2013,Li2017}.
	Table~\ref{tab:lbrk} compares the jet breakup lengths from Ref.~\citep{Matsuo2013} and with those in our simulations; the error $E$ here is defined as 
\begin{equation}
	E = \frac{|L_{\mathrm{exp}} - L_{\mathrm{sim}}|}{L_{\mathrm{exp}}}\times100,
\end{equation}
where $L_{\mathrm{exp}}$ and $L_{\mathrm{sim}}$ are the breakup lengths obtained via the experiments and simulations, respectively.	
	The simulation's accuracy improves as $D_{j0}$ increases, the error dropping from 31.4\% for Case 1 to 10.8\% for Case 3.
	Thus, these simulations were able to predict the experimental jet breakup length $L$ to within a maximum error of 31.4\%.

\begingroup
\renewcommand{\arraystretch}{1.3}
\begin{table}[tb]
\caption{Comparison of the jet breakup lengths observed in the experiments and simulations.
	The simulations predicted the experimental breakup length $L$ to within a maximum error of 31.4\%.
\label{tab:lbrk} }
\begin{ruledtabular}
\begin{tabular}{cccc}
 & Experiment~\citep{Matsuo2013} [mm] & Simulation [mm] & Error $E$ (\%)   \\
\colrule
Case 1 & $171$ & $117$ & $31.4$ \\
Case 2 & $264$ & $337$ & $27.8$ \\
Case 3 & $348$ & $310$ & $10.8$ \\
Case 4 & $440$ & $382$ & $13.0$ \\
\end{tabular}
\end{ruledtabular}
\end{table}
\endgroup

\begin{figure}[tb]
\begin{center}
	\includegraphics[width=8.5cm]{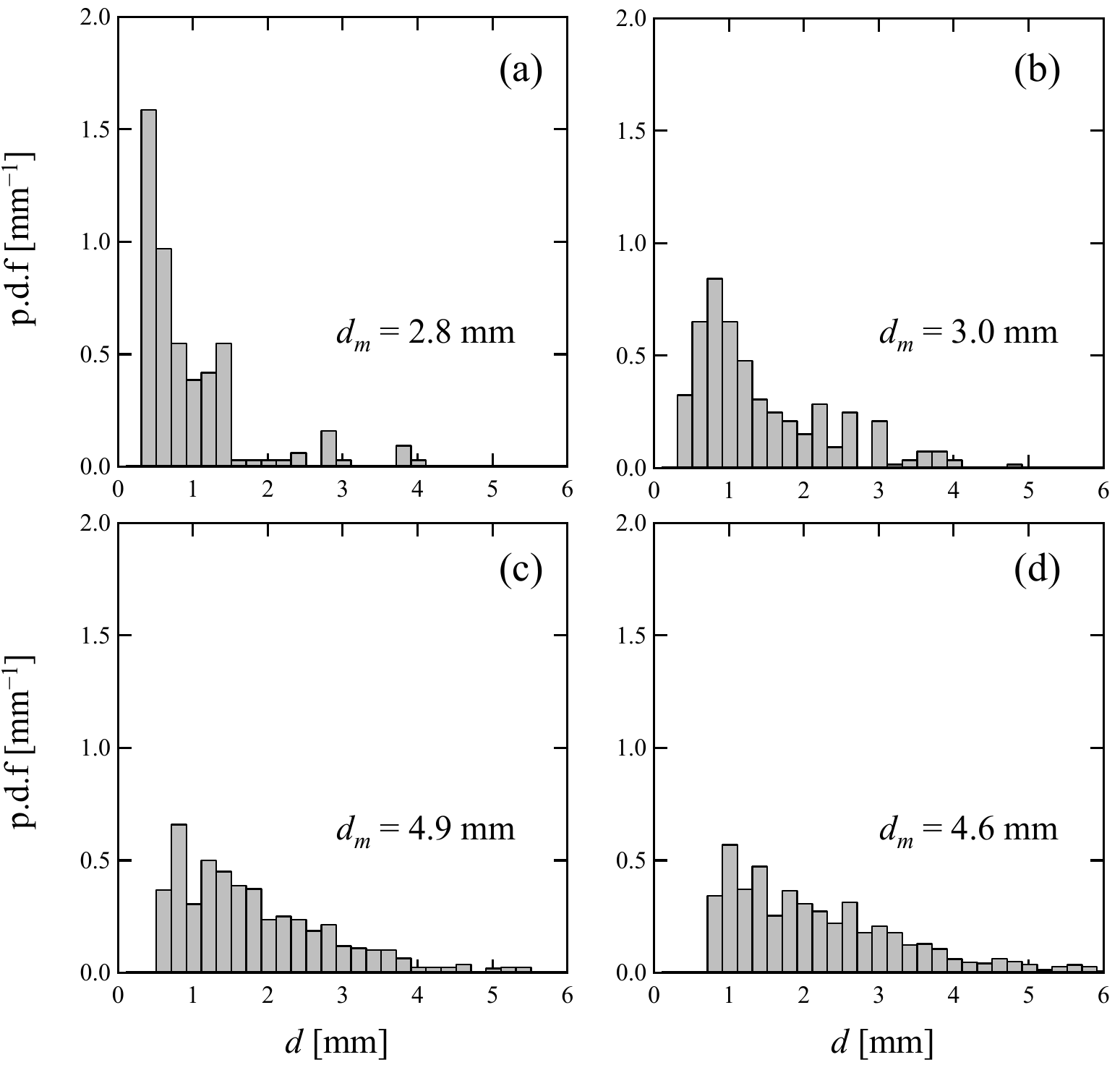}
	\caption{Fragment-diameter distribution histograms for (a) Case 1 ($D_{j0}=7$ mm), (b) Case 2 ($D_{j0}=10$ mm), (c) Case 3 ($D_{j0}=15$ mm), and (d) Case 4 ($D_{j0}=20$ mm).
	Here, $d_m$ denotes the mass-median diameter.
	Larger inlet diameter $D_{j0}$ lead to large fragments appearing more frequently.
	\label{fig:diameterDist}}
\end{center}
\end{figure}

	As another quantitative comparison, we also evaluated the fragment diameters.
	The fragment diameter $d$ was determined by the following procedures:
\begin{enumerate}
	\item Binarize the order parameter $\phi$ into 1 for the jet region and 0 for the others
	\item Find the connected components (with 1) and regard them as fragments
	\item Calculate each fragment's volume $V$
	\item Convert the volume $V$ into the equivalent spherical diameter by $d \equiv (6V/\pi)^{1/3}$
\end{enumerate}
	Figure~\ref{fig:diameterDist} shows histograms of the measured diameters; we have excluded the continuous liquid column from the nozzle from the calculation.
	From the figure, we can find that the smaller the nozzle diameter is, the higher the observation frequency of the droplet is.
	Except for $D_{j0} = 7$ mm [Fig.~\ref{fig:diameterDist}(a)], all the distributions have   a long tail to the right, similar to a log-normal distribution.
	
	Next, we compared the fragment diameters measured via the simulations with the experimental data and the predictions of hydrodynamic instability theories. 
	We calculated the mass-median diameter as a fragment size metric, due to the shape of the distribution (Fig.~\ref{fig:diameterDist}).
	On the theoretical side, the first indicators we compared were the critical wavelengths of classical Rayleigh--Taylor (RT) and Kelvin--Helmholtz (KH) instabilities, $\lambda_{\mathrm{cr,RT}}$ and $\lambda_{\mathrm{cr,KH}}$.
	Assuming a two-dimensional stratified geometry, these are~\cite{Chandrasekhar1961}
\begin{align}
	\lambda_{\mathrm{cr,RT}} &= 2\pi \left[\frac{\sigma}{(\rho_j - \rho_c)g} \right]^{\frac{1}{2}}, \label{eq:RT} 
	\\
	\lambda_{\mathrm{cr,KH}} &= 2\pi \frac{\rho_j+\rho_c}{\rho_j \rho_c} \frac{\sigma}{u_{j0}^2}.  \label{eq:KH}
\end{align}
	Here, we used the jet velocity $u_{j0}$ as the scaling velocity, assuming that the ambient fluid was stationary in Eq.~(\ref{eq:KH}).
	The second theoretical indicator we compared was the critical Weber number $\mathrm{We}_\mathrm{cr}$, from which we obtained the critical droplet diameter $d$, as follows:
\begin{equation}
	d = \mathrm{We}_{\mathrm{cr}}\cdot \frac{\sigma}{\rho_c u_{j0}^2}. \label{eq:CritWe}
\end{equation}
	The value of $\mathrm{We}_\mathrm{cr}$ depends on the assumptions made, so several values have been proposed, such as 12 (\citet{Pilch1987}) and 18 (\citet{Matsuo2008}).
	Previous studies~\citep{Bang2003,Abe2006,Li2017} have pointed out that these theoretical quantities are related to the sizes of the fragments generated by jet breakup.
	
\begin{figure}[tb]
\begin{center}
	\includegraphics[width=7.0cm]{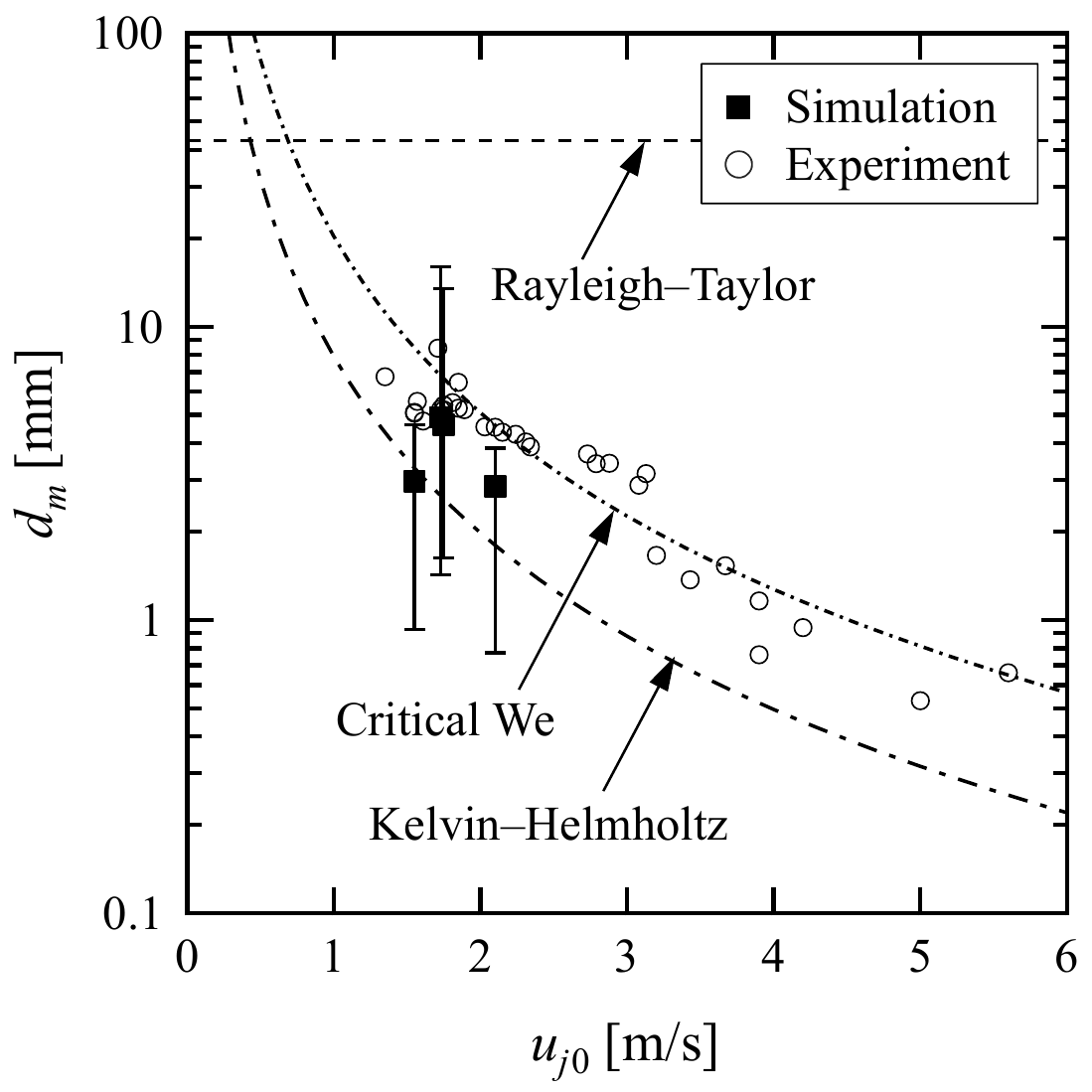}
	\caption{Comparison of the experimental and simulated fragment sizes. 
	Here, the critical Weber number $\mathrm{We}_{\mathrm{cr}}$ in Eq.~(\ref{eq:CritWe}) was assumed to be 18.
	Both the experimental~\citep{Matsuo2013} and simulation results are in accordance with the Kelvin--Helmholtz [Eq.~(\ref{eq:KH})] and critical Weber number  [Eq.~(\ref{eq:CritWe})] theories.
	However, the Rayleigh--Taylor instability [Eq.~(\ref{eq:RT})] does not correlate with either the numerical or experimental results.
	\label{fig:fragSize}}
\end{center}
\end{figure}

	Figure~\ref{fig:fragSize} compares the simulated and experimental fragment sizes.
	The aforementioned hydrodynamic instability theories [Eqs.~(\ref{eq:RT})--(\ref{eq:CritWe})] are also shown in the graph.
	The error bars indicate the 95\% confidence intervals for the median diameter, which we used to express the widths of the droplet size distributions even though they were not necessarily Gaussian~\citep{Saito2017}.
	As in the experiment~\citep{Matsuo2013}, the simulation results are in agreement with both the KH [Eq.~(\ref{eq:KH})] and critical Weber number [Eq.~(\ref{eq:CritWe})] theories.
	However, the RT instability [Eq.~(\ref{eq:RT})] does not appear to be correlated with the numerical or experimental results.
	As in Figs.~(\ref{fig:jet_d07})--(\ref{fig:jet_d20}), the jet's interfaces appear to be very unstable at the jet-side region both in the experiments and simulations.
	Since there is velocity difference between the jet and the ambient fluid, KH instability may be one of the reasons of the onset of interfacial instability.
	In addition, fragments generated at the tip of the jet (also at the jet side in some cases) are considered to be fragmented into smaller drops again due to the limitation of the critical Weber number.
	These simulations thus support the breakup mechanism proposed based on the experimental observations.
	Table~\ref{tab:median} compares the experimental and simulated median diameters; the error $E$ is defined as
\begin{equation}
	E = \frac{|d_{m,\mathrm{exp}} - d_{m,\mathrm{sim}}|}{d_{m,\mathrm{exp}}}\times100,
\end{equation}	
	where $d_{m,\mathrm{exp}}$ and $d_{m,\mathrm{sim}}$ are the experimental and simulated median diameters, respectively.
	The maximum error $E$ was 41.9\%, for Case 2.
	
\begingroup
\renewcommand{\arraystretch}{1.3}
\begin{table}[tb]
\caption{Comparison of the mass-median diameters observed in the experiments and simulations.
	The simulations predicted the experimental median diameter $d_m$ to within a maximum error of 41.9\%.
\label{tab:median} }
\begin{ruledtabular}
\begin{tabular}{cccc}
 & Experiment~\citep{Matsuo2013} [mm] & Simulation [mm] & Error $E$ (\%)   \\
\colrule
Case 1 & $4.54$ & $2.85$ & $37.2$ \\
Case 2 & $5.10$ & $2.96$ & $41.9$ \\
Case 3 & $5.23$ & $4.91$ & $6.0 $ \\
Case 4 & $5.40$ & $4.61$ & $14.4$ \\
\end{tabular}
\end{ruledtabular}
\end{table}
\endgroup

\subsection{FT experiments \label{sec:FT}}

	The FT experiments were performed at the Joint Research Centre (JRC) in Ispra (Italy) by \citet{Magallon1992}.
	Around 100-kg-scale of UO$_2$ melt was poured into a liquid-sodium pool.
	Two experiments, called T1 and T2, were carried out, with release diameters of 50 mm and 80 mm, respectively. 
	In this paper, we focus on the T1 experiment.
	A notable feature of the FT experiments is that sodium in not transparent, so these simulations will hopefully help us to better understand the phenomena involved.

\begin{figure}[tb]
\begin{center}
	\includegraphics[width=6.8cm]{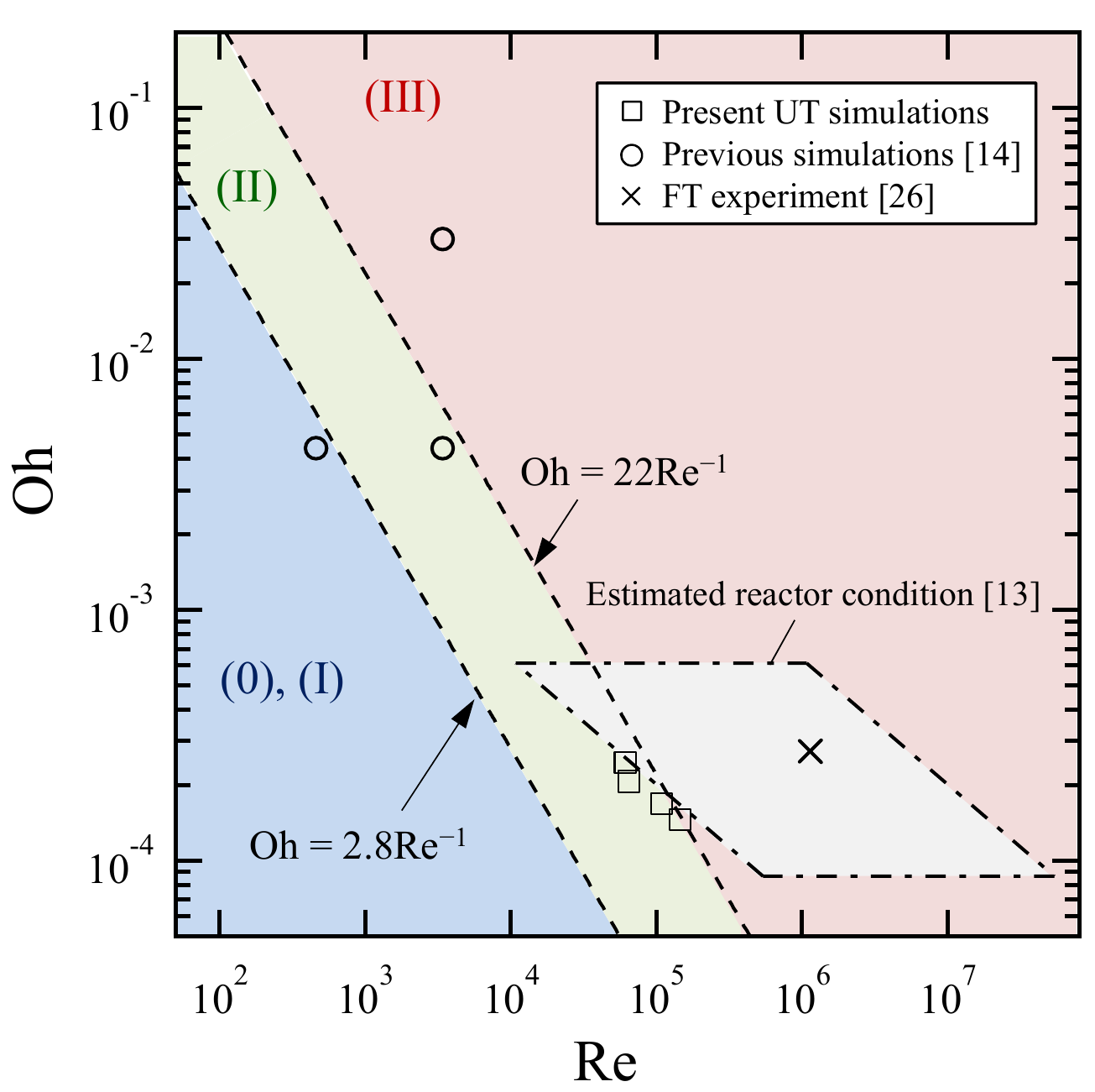}
	\caption{Flow regime region covered by the FT experiment together with the conditions our previous simulations (`o')~\citep{Saito2017b}. 
	Our current simulations (`$\times$') involve substantially higher Re values compared with the previous ones.The present simulation condition  is located extremely higher $\mathrm{Re}$ condition compared with our previous simulation.
	The hatched gray region shows the reactor conditions estimated in Ref.~\citep{Saito2017};
	the FT experiment fell within this region.
	This diagram predicts that the breakup will be in the atomization regime (III).
	The conditions of UT simulations presented in Sec.~\ref{sec:UT} are also shown in the same map for reference.
	\label{fig:FTcondition}}
\end{center}
\end{figure}

	For these simulations, we discretized the computational domain into an $8D_{j0}\times8D_{j0}\times20D_{j0}$ lattice.
	Table~\ref{tab:FT} summarizes the simulation conditions, together with the corresponding dimensionless quantities, given by Eqs.~(\ref{eq:densityRatio})--(\ref{eq:Froude}). 
	It should be noted that the Reynolds number in Table~\ref{tab:FT} ($1.1\times10^6$) is extremely high from a multiphase LB perspective. 
	The liquid--liquid-jet breakup flow-regime map~\cite{Saito2017} shown in Fig.~\ref{fig:FTcondition} indicates we are in the atomization regime (Regime III) in this case. 
	To examine the effect of grid resolution on jet breakup behavior, we considered two cases with different grids: a {\it coarse} case with $D_{j0} = 30$ and hence a  $240\times240\times600$ lattice $(=34\,560\,000~{\rm grid~points})$,
	and a {\it fine} case with $D_{j0} = 60$ and hence a $480\times480\times1\,200$ lattice $(=276\,480\,000~{\rm grid~points})$.
   The resulting minimum spacings (in physical units) were $\Delta x = 1.67$ mm and $\Delta x = 0.83$ mm for the coarse and fine cases, respectively.
   For reference, the conditions of UT simulations presented in Sec.~\ref{sec:UT} are also shown in the same map.
	
\begingroup
\renewcommand{\arraystretch}{1.4}
\begin{table*}[tb]
\caption{\label{tab:FT} 
Conditions for the FT experiment simulations, reproduced from Ref.~\citep{Magallon1992}.
Here, we focus on their T1 experiment.
	The physical properties for the UO$_2$ melt and liquid sodium were as follows~\citep{Chawla1981}: $\rho_j=8\,663~\mathrm{kg/m^3}$, $\nu_j=0.46~\mathrm{mm^2/s}$, $\sigma=0.465~\mathrm{N/m}$, $\rho_c=856~\mathrm{kg/m^3}$, and $\nu_c=0.28~\mathrm{mm^2/s}$.
	The dimensionless quantities, calculated using Eqs.~(\ref{eq:densityRatio})--(\ref{eq:Froude}), are also shown,
including the density ratio $\gamma$, kinematic viscosity ratio $\eta$, Reynolds number $\mathrm{Re}$, Weber number $\mathrm{We}$, and Froude number $\mathrm{Fr}$.}
\begin{ruledtabular}
\begin{tabular}{cccccccc}
 $D_{j0}~[\mathrm{mm}]$ & 
 $u_{j0}~[\mathrm{m/s}]$ & 
 $\gamma~[-]$ & 
 $\eta~[-]$ & 
 $\mathrm{Re}~[-]$ & 
 $\mathrm{We}~[-]$ & 
 $\mathrm{Fr}~[-]$ \\
\colrule
 $50$ & $10$ & $10.1$ & $ 1.4 $ & $1.1\times10^6$ & $9.3\times10^4$ & $2.0\times10^2$\\
\end{tabular}
\end{ruledtabular}
\end{table*}
\endgroup

\begin{figure*}[tb]
\begin{center}
	\includegraphics[width=13cm]{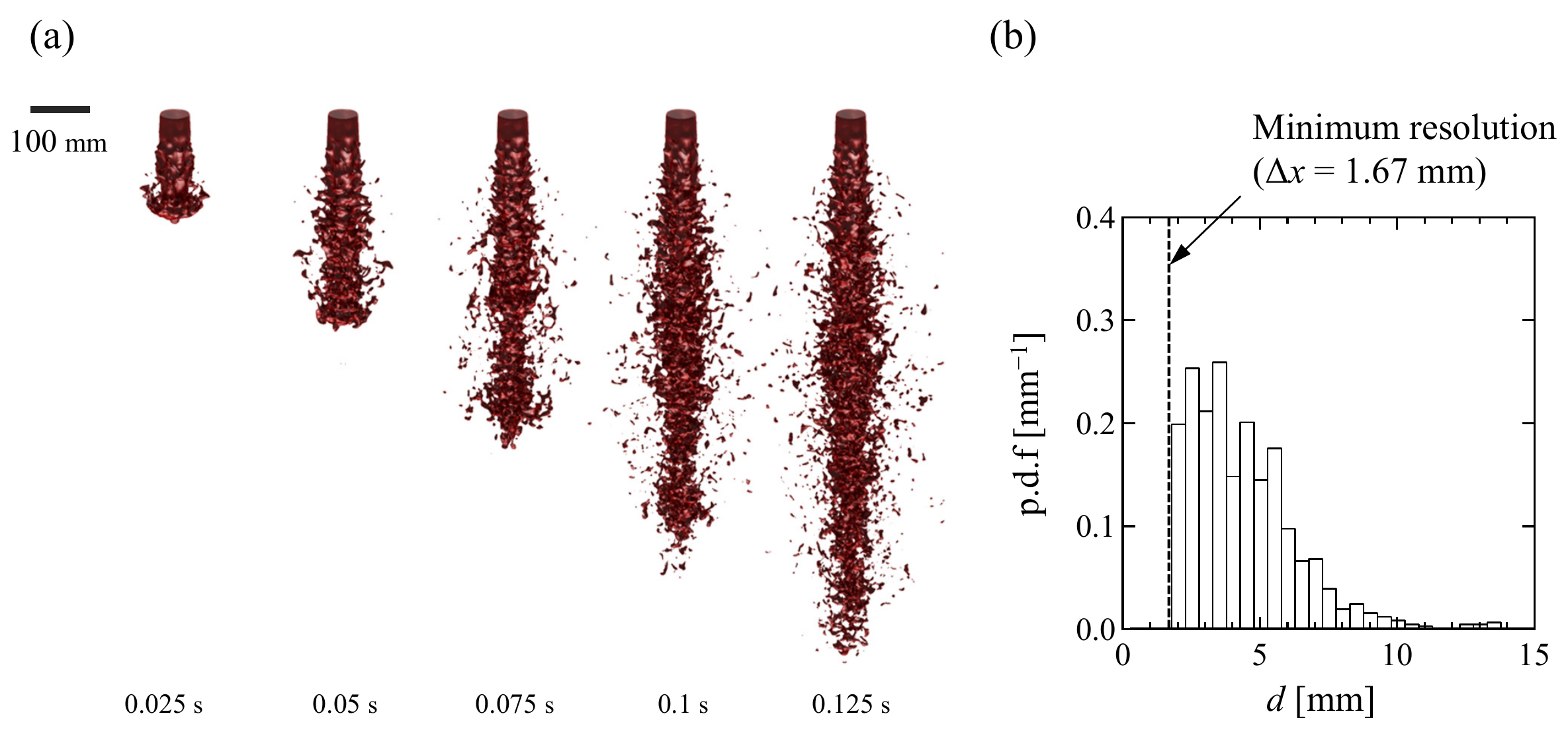}
	\caption{FT simulation results for the coarse (low-resolution) case, showing the (a) interfacial behavior and (b) fragment-diameter distribution.
	The inlet diameter was set to $D_{j0}=30$ (in lattice units),
	and the computational domain was discretized into a $240\times240\times600$  lattice $(=34\,560\,000~{\rm grid}~{\rm points})$.
	The simulation parameters were as follows (in lattice units): $\sigma=1.6\times10^{-5}$, $\nu_j=\nu_r=1.4\times10^{-6}$, $\nu_c=\nu_b=9.9\times10^{-7}$, $g=4.1\times10^{-7}$, $\rho_j=\rho_r^0=1$, and $\rho_c=\rho_b^0=0.099$.
	The minimum spatial resolution in this case was $\Delta x = 1.67$ mm.
	Although the computation is stable, some unphysical behavior can be seen: the fragments do not appear to be spherical, and the fragments that should appear around the jet's leading edge are not present. 
	In addition, the fragment-diameter distribution appears to be log-normal but is interrupted near the minimum resolution $\Delta x$.
	\label{fig:T1_1}}
\end{center}
\end{figure*}

	Figure~\ref{fig:T1_1} shows the simulation results for the coarse case.
	The left hand side [Fig.~\ref{fig:T1_1}(a)] shows the time evolution of the jet interface in real units, 
   while the right hand side [Fig.~\ref{fig:T1_1}(b)] shows the calculated fragment size distribution at $t = 0.125$ s. 
   The simulation parameters are given (in lattice units) in the caption.
	Even though the numerical conditions were somewhat extreme, numerical stability was maintained throughout the simulations 
	These results show that using our CG LB model, based on nonorthogonal CMs, allowed $\mathrm Re$ to be increased significantly. 
	In Fig.~\ref{fig:T1_1}, most of the fragments are generated at the side of the jet due to entrainment, and they appear to be tiny compared with the inlet diameter $D_{j0}$.
	This is characteristic of the atomization regime (Regime III)~\citep{Saito2017}.
	This simulation therefore suggests that the jet state in the FT experiment should be similar to the atomization regime and, in fact, the debris collected in the T1 experiment was very fine (around 30--600 $\mathrm{\mu}$m)~\citep{Magallon1992}.
	That said, however, some unphysical points remain in terms of the fragment shapes and numerical dissolution.
	The resulting fragments were not spherical but irregular, and the jet's leading edge faced numerical dissolution at the later stage compared to that at the initial stages.
	In addition, Fig.~\ref{fig:T1_1}(b) shows that the fragment-diameter distribution appears to be log-normal, but this pattern is interrupted near the minimum resolution, which we believe is due to the low spatial resolution.

\begin{figure*}[tb]
\begin{center}
	\includegraphics[width=13cm]{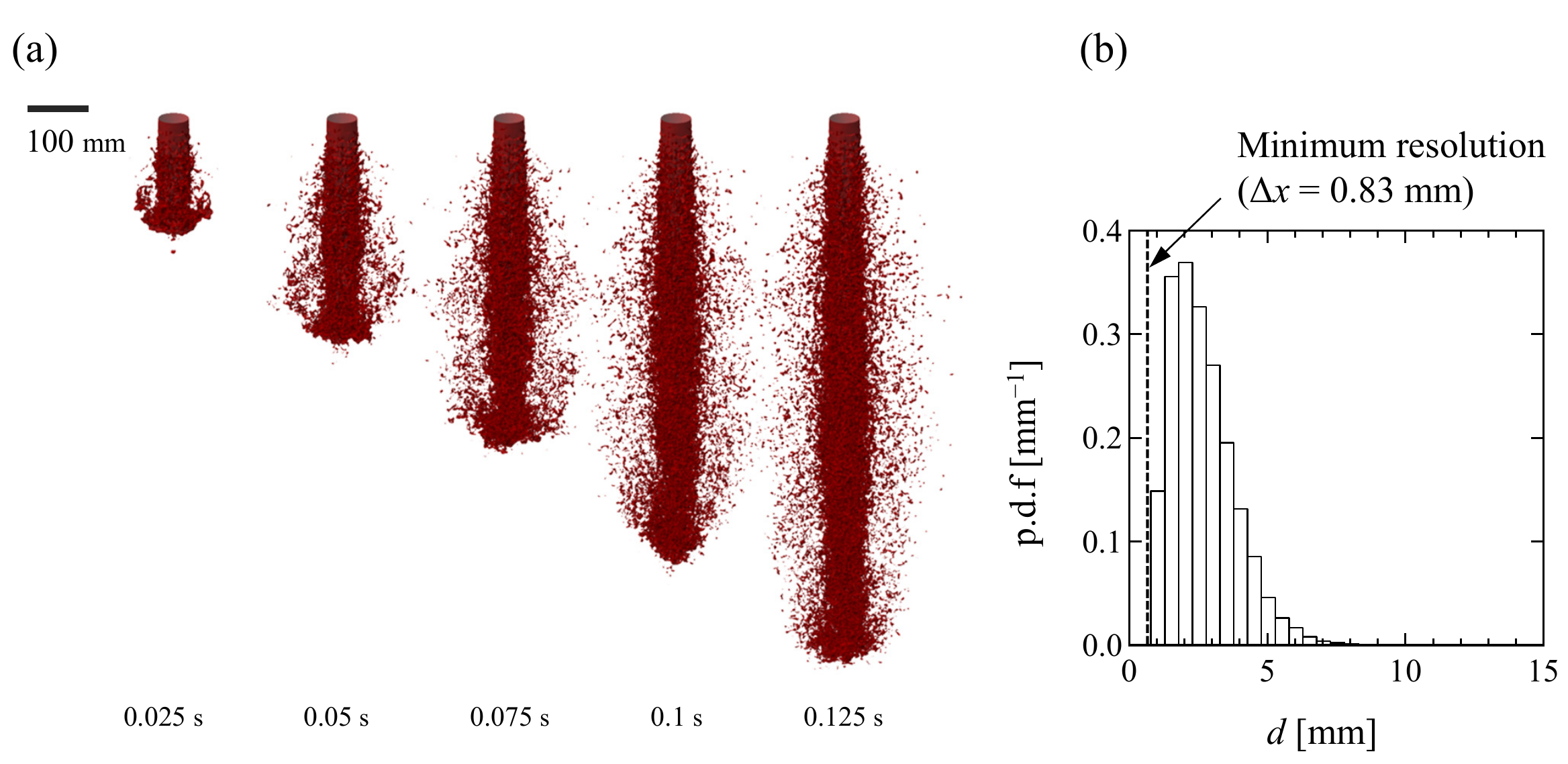}
	\caption{FT simulation results for the fine (high-resolution) case, showing the (a) interfacial behavior and (b) fragment diameter distribution.
	The inlet diameter was set to $D_{j0}=60$ (in lattice units), and the computational domain is discretized into a $480\times480\times1\,200$ lattice $(=276\,480\,000~{\rm grid}~{\rm points})$.
	The simulation parameters used were as follows (in lattice units): $\sigma=3.2\times10^{-5}$, $\nu_j=\nu_r=2.8\times10^{-6}$, $\nu_c=\nu_b=2.0\times10^{-6}$, $g=2.0\times10^{-7}$, $\rho_j=\rho_r^0=1$, and $\rho_c=\rho_b^0=0.099$.
	The minimum spatial resolution in this case was $\Delta x = 0.83$ mm.
	Compared with the results in Fig.~\ref{fig:T1_1}(a), the interfacial behavior seen in Fig.~\ref{fig:T1_2}(a) is clearly different in terms of the fragment sizes and shapes: 
	the fragments are smaller and nearly spherical.
	In addition, the fragment-diameter distribution in Fig.~\ref{fig:T1_2}(b) is smoother than that in Fig.~\ref{fig:T1_1}(b), and the higher-resolution domain has also improved 
the numerical dissolution.
	\label{fig:T1_2}}
\end{center}
\end{figure*}

	To investigate the effect of spatial resolution, we then carried out a fine simulation, in which there were twice as many grid points in each direction. 
	We used the same dimensionless quantities (Table~\ref{tab:FT}) as in the coarse case.
	Figure~\ref{fig:T1_2} shows the results of this simulation, with the simulation parameters again shown (in lattice units) in the caption. 
	Compared with Fig.~\ref{fig:T1_1}(a), the interfacial behavior seen in Fig.~\ref{fig:T1_2}(a) is clearly different in terms of the fragment sizes and shapes: they are smaller and nearly spherical, and fragmentation now occurs around the jet's leading edge and side. 
	The fragment-diameter distribution in Fig.~\ref{fig:T1_2}(b) is also smoother than that in Fig.~\ref{fig:T1_1}(b), resembling a log-normal distribution. 
    Adopting a higher-resolution computational domain also improved the numerical dissolution issue, as more physically reasonable results were obtained.
    In both cases (coarse and fine), the atomization regime appeared, implying that the grid resolutions used in this paper were sufficient to simulate the qualitative behavior of the entire jet. 
    However, we can also observe that finer grids will be required to perform quantitative evaluations.

\section{Conclusion\label{sec:conclusions}}

	In this paper, we have extended the previous CG LB model~\citep{Saito2017b} by introducing nonorthogonal CMs in three dimensions~\cite{DeRosis2017a} within the GMRT framework~\citep{Fei2017, Fei2018b}.
	Static droplet tests showed that our model can predict the interfacial tension for a range of density ratios (up to $1\,000$) to within a maximum error of 0.40\% and also that it can greatly reduce the spurious velocity.
	We also applied our model to hydrodynamic melt-jet breakup simulations, targeting two different experiments, namely, the UT~\cite{Matsuo2013} and FT~\cite{Magallon1992} experiments.
	Numerical simulations under corresponding conditions, including those equivalent to an actual reactor, demonstrated that our model was more stable.
	In particular, our model allowed the Reynolds number to be increased significantly, up to $O(10^6)$. 
	The UT simulations predicted the jet breakup length and median fragment-diameter to within maximum errors of 31.4\% and 41.9\%, respectively.
	The results of the FT simulation suggested that the jet in the experiment was in the atomization regime.
	To investigate the effect of grid resolution, the latter simulation was carried out at two grid resolutions, with a minimum spacing of $\Delta x=0.83$ mm. 
	The results imply that finer grid resolutions will be required to evaluate the behavior accurately.
	Since the spatial resolution used was limited by our computing environment, we plan to use a higher-performance environment in our future work.

%
%

\begin{acknowledgments}
	This work was supported by Mitsubishi Heavy Industries, Ltd. 
	The authors are grateful to H. Sakaba and H. Sato for their invaluable support. 
	The support of JSPS KAKENHI Grant Number 16J02077 is also acknowledged.
	This research used computational resources of COMA in Center for Computational Sciences, University of Tsukuba, and Earth Simulator in JAMSTEC.
	This article is based upon work from COST Action MP1305, supported by COST (European Cooperation in Science and Technology).
\end{acknowledgments}

\clearpage

\begin{widetext}
\appendix
\section{\label{sec:appA}Transformation and shift matrices}
When the lattice velocity ${\bf c}_i$ is defined by Eq.~(\ref{eq:velVec}), the transformation matrix {\bf M} can be given by
\begin{equation}
\scalebox{0.8}{$\displaystyle
{\bf M} = 
\left [
	\begin{array}{ccccccccccccccccccccccccccc}
		1 & 1 & 1 & 1 & 1 & 1 & 1 & 1 & 1 & 1 & 1 & 1 & 1 & 1 & 1 & 1 & 1 & 1 & 1 & 1 & 1 & 1 & 1 & 1 & 1 & 1 & 1 \\	
		0	&1	&-1	&0	&0	&0	&0	&1	&-1	&1	&-1	&0	&0	&0	&0	&1	&-1	&1	&-1	&1	&-1	&1	&-1	&1	&-1	&-1	&  1 \\	
		0	&0	&0	&1	&-1	&0	&0	&1	&-1	&-1	&1	&1	&-1	&1	&-1	&0	&0	&0	&0	&1	&-1	&1	&-1	&-1	&1	&1	& -1 \\	
		0	&0	&0	&0	&0	&1	&-1	&0	&0	&0	&0	&1	&-1	&-1	&1	&1	&-1	&-1	&1	&1	&-1	&-1	&1	&1	&-1	&1	& -1 \\	
		0	&0	&0	&0	&0	&0	&0	&1	&1	&-1	&-1	&0	&0	&0	&0	&0	&0	&0	&0	&1	&1	&1	&1	&-1	&-1	&-1	& -1 \\	
		0	&0	&0	&0	&0	&0	&0	&0	&0	&0	&0	&0	&0	&0	&0	&1	&1	&-1	&-1	&1	&1	&-1	&-1	&1	&1	&-1	& -1 \\	
		0	&0	&0	&0	&0	&0	&0	&0	&0	&0	&0	&1	&1	&-1	&-1	&0	&0	&0	&0	&1	&1	&-1	&-1	&-1	&-1	&1	&  1 \\	
		0	&1	&1	&-1	&-1	&0	&0	&0	&0	&0	&0	&-1	&-1	&-1	&-1	&1	&1	&1	&1	&0	&0	&0	&0	&0	&0	&0	&  0 \\	
		0	&1	&1	&0	&0	&-1	&-1	&1	&1	&1	&1	&-1	&-1	&-1	&-1	&0	&0	&0	&0	&0	&0	&0	&0	&0	&0	&0	&  0 \\	
		0	&1	&1	&1	&1	&1	&1	&2	&2	&2	&2	&2	&2	&2	&2	&2	&2	&2	&2	&3	&3	&3	&3	&3	&3	&3	&  3 \\	
		0	&0	&0	&0	&0	&0	&0	&1	&-1	&1	&-1	&0	&0	&0	&0	&1	&-1	&1	&-1	&2	&-2	&2	&-2	&2	&-2	&-2	&  2 \\	
		0	&0	&0	&0	&0	&0	&0	&1	&-1	&-1	&1	&1	&-1	&1	&-1	&0	&0	&0	&0	&2	&-2	&2	&-2	&-2	&2	&2	& -2 \\	
		0	&0	&0	&0	&0	&0	&0	&0	&0	&0	&0	&1	&-1	&-1	&1	&1	&-1	&-1	&1	&2	&-2	&-2	&2	&2	&-2	&2	& -2 \\	
		0	&0	&0	&0	&0	&0	&0	&1	&-1	&1	&-1	&0	&0	&0	&0	&-1	&1	&-1	&1	&0	&0	&0	&0	&0	&0	&0	&  0 \\	
		0	&0	&0	&0	&0	&0	&0	&1	&-1	&-1	&1	&-1	&1	&-1	&1	&0	&0	&0	&0	&0	&0	&0	&0	&0	&0	&0	&  0 \\	
		0	&0	&0	&0	&0	&0	&0	&0	&0	&0	&0	&-1	&1	&1	&-1	&1	&-1	&-1	&1	&0	&0	&0	&0	&0	&0	&0	&  0 \\	
		0	&0	&0	&0	&0	&0	&0	&0	&0	&0	&0	&0	&0	&0	&0	&0	&0	&0	&0	&1	&-1	&-1	&1	&-1	&1	&-1	&  1 \\	
		0	&0	&0	&0	&0	&0	&0	&1	&1	&1	&1	&1	&1	&1	&1	&1	&1	&1	&1	&3	&3	&3	&3	&3	&3	&3	&  3 \\	
		0	&0	&0	&0	&0	&0	&0	&1	&1	&1	&1	&-1	&-1	&-1	&-1	&1	&1	&1	&1	&1	&1	&1	&1	&1	&1	&1	&  1 \\	
		0	&0	&0	&0	&0	&0	&0	&1	&1	&1	&1	&0	&0	&0	&0	&-1	&-1	&-1	&-1	&0	&0	&0	&0	&0	&0	&0	&  0 \\	
		0	&0	&0	&0	&0	&0	&0	&0	&0	&0	&0	&0	&0	&0	&0	&0	&0	&0	&0	&1	&1	&-1	&-1	&-1	&-1	&1	&  1 \\	
		0	&0	&0	&0	&0	&0	&0	&0	&0	&0	&0	&0	&0	&0	&0	&0	&0	&0	&0	&1	&1	&-1	&-1&	1	&1	&-1	 &-1 \\	
		0	&0	&0	&0	&0	&0	&0	&0	&0	&0	&0	&0	&0	&0	&0	&0	&0	&0	&0	&1	&1	&1	&1	&-1	&-1	&-1	 &-1 \\	
		0	&0	&0	&0	&0	&0	&0	&0	&0	&0	&0	&0	&0	&0	&0	&0	&0	&0	&0	&1	&-1	&1	&-1	&1	&-1	&-1	 & 1 \\	
		0	&0	&0	&0	&0	&0	&0	&0	&0	&0	&0	&0	&0	&0	&0	&0	&0	&0	&0	&1	&-1	&1	&-1	&-1	&1	&1	 &-1 \\	
		0	&0	&0	&0	&0	&0	&0	&0	&0	&0	&0	&0	&0	&0	&0	&0	&0	&0	&0	&1	&-1	&-1	&1	&1	&-1	&1	 &-1 \\	
		0	&0	&0	&0	&0	&0	&0	&0	&0	&0	&0	&0	&0	&0	&0	&0	&0	&0	&0	&1	&1	&1	&1	&1	&1	&1	 & 1 \\	
	\end{array}
	\right] $}, \label{eq:M}
\end{equation}
and its inverse ${\bf M}^{-1}$ can be given by
\begin{equation}
\scalebox{0.75}{$\displaystyle
{\bf M}^{-1} = \frac{1}{48}
\left [
	\begin{array}{ccccccccccccccccccccccccccc}
		 48	&0	&0	&0	&0	&0	&0	&0	&0	&-48	&0	&0	&0	&0	&0	&0	&0	&48	&0	&0	&0	&0	&0	&0	&0	&0	 &-48\\	
		 0	&24	&0	&0	&0	&0	&0	&8	&8	&8	&-24	&0	&0	&0	&0	&0	&0	&-12	&-12	&0	&0	&0	&0	&24	&0	&0	&  24\\	
		 0	&-24	&0	&0	&0	&0	&0	&8	&8	&8	&24	&0	&0	&0	&0	&0	&0	&-12	&-12	&0	&0	&0	&0	&-24	&0	&0	&  24\\	
		 0	&0	&24	&0	&0	&0	&0	&-16	&8	&8	&0	&-24	&0	&0	&0	&0	&0	&-18	&6	&-12	&0	&0	&0	&0	&24	&0	&  24\\	
		 0	&0	&-24	&0	&0	&0	&0	&-16	&8	&8	&0	&24	&0	&0	&0	&0	&0	&-18	&6	&-12	&0	&0	&0	&0	&-24	&0	 & 24\\	
		 0	&0	&0	&24	&0	&0	&0	&8	&-16	&8	&0	&0	&-24	&0	&0	&0	&0	&-18	&6	&12	&0	&0	&0	&0	&0	&24	&  24\\	
		 0	&0	&0	&-24	&0	&0	&0	&8	&-16	&8	&0	&0	&24	&0	&0	&0	&0	&-18	&6	&12	&0	&0	&0	&0	&0	&-24	&  24\\	
		 0	&0	&0	&0	&12	&0	&0	&0	&0	&0	&6	&6	&0	&6	&6	&0	&0	&3	&3	&6	&0	&0	&-12	&-12	&-12	&0	 &-12\\	
		 0	&0	&0	&0	&12	&0	&0	&0	&0	&0	&-6	&-6	&0	&-6	&-6	&0	&0	&3	&3	&6	&0	&0	&-12	&12	&12	&0	 &-12\\	
		 0	&0	&0	&0	&-12	&0	&0	&0	&0	&0	&6	&-6	&0	&6	&-6	&0	&0	&3	&3	&6	&0	&0	&12	&-12	&12	&0	 &-12\\	
		 0	&0	&0	&0	&-12	&0	&0	&0	&0	&0	&-6	&6	&0	&-6	&6	&0	&0	&3	&3	&6	&0	&0	&12	&12	&-12	&0	& -12\\	
		 0	&0	&0	&0	&0	&0	&12	&0	&0	&0	&0	&6	&6	&0	&-6	&-6	&0	&6	&-6	&0	&-12	&0	&0	&0	&-12	&-12	& -12\\	
		 0	&0	&0	&0	&0	&0	&12	&0	&0	&0	&0	&-6	&-6	&0	&6	&6	&0	&6	&-6	&0	&-12	&0	&0	&0	&12	&12	& -12\\	
		 0	&0	&0	&0	&0	&0	&-12	&0	&0	&0	&0	&6	&-6	&0	&-6	&6	&0	&6	&-6	&0	&12	&0	&0	&0	&-12	&12	& -12\\	
		 0	&0	&0	&0	&0	&0	&-12	&0	&0	&0	&0	&-6	&6	&0	&6	&-6	&0	&6	&-6	&0	&12	&0	&0	&0	&12	&-12	& -12\\	
		 0	&0	&0	&0	&0	&12	&0	&0	&0	&0	&6	&0	&6	&-6	&0	&6	&0	&3	&3	&-6	&0	&-12	&0	&-12	&0	&-12	& -12\\	
		 0	&0	&0	&0	&0	&12	&0	&0	&0	&0	&-6	&0	&-6&	6	&0	&-6	&0	&3	&3	&-6	&0	&-12	&0	&12	&0	&12	& -12\\	
		 0	&0	&0	&0	&0	&-12	&0	&0	&0	&0	&6	&0	&-6	&-6	&0	&-6	&0	&3	&3	&-6	&0	&12	&0	&-12	&0	&12	& -12\\	
		 0	&0	&0	&0	&0	&-12	&0	&0	&0	&0	&-6	&0	&6	&6	&0	&6	&0	&3	&3	&-6	&0	&12	&0	&12	&0	&-12	& -12\\	
		 0	&0	&0	&0	&0	&0	&0	&0	&0	&0	&0	&0	&0	&0	&0	&0	&6	&0	&0	&0	&6	&6	&6	&6	&6	&6	&   6\\	
		 0	&0	&0	&0	&0	&0	&0	&0	&0	&0	&0	&0	&0	&0	&0	&0	&-6	&0	&0	&0	&6	&6	&6	&-6	&-6	&-6	&   6\\	
		 0	&0	&0	&0	&0	&0	&0	&0	&0	&0	&0	&0	&0	&0	&0	&0	&-6	&0	&0	&0	&-6	&-6	&6	&6	&6	&-6	&   6\\	
		 0	&0	&0	&0	&0	&0	&0	&0	&0	&0	&0	&0	&0	&0	&0	&0	&6	&0	&0	&0	&-6	&-6	&6	&-6	&-6	&6	&   6\\	
		 0	&0	&0	&0	&0	&0	&0	&0	&0	&0	&0	&0	&0	&0	&0	&0	&-6	&0	&0	&0	&-6	&6	&-6	&6	&-6	&6	&   6\\	
		 0	&0	&0	&0	&0	&0	&0	&0	&0	&0	&0	&0	&0	&0	&0	&0	&6	&0	&0	&0	&-6	&6	&-6	&-6	&6	&-6	&   6\\	
		 0	&0	&0	&0	&0	&0	&0	&0	&0	&0	&0	&0	&0	&0	&0	&0	&-6	&0	&0	&0	&6	&-6	&-6	&-6	&6	&6	&   6\\	
		 0	&0	&0	&0	&0	&0	&0	&0	&0	&0	&0	&0	&0	&0	&0	&0	&6	&0	&0	&0	&6	&-6	&-6	&6	&-6	&-6	&   6\\	
	\end{array}
	\right] $}.
\end{equation}
	In addition, the shift matrix ${\bf N}$ can be given by
\begin{align}
\thickmuskip=0mu
\medmuskip=0mu
\thinmuskip=0mu
\rotatebox{90}{
\scalebox{0.43}{$\displaystyle
{\bf N} = 
\left [
	\begin{array}{ccccccccccccccccccccccccccc}
		1	&0&	0&	0&	0&	0&	0&	0&	0&	0&	0&	0&	0&	0&	0&	0&	0&	0&	0&	0&	0&	0&	0&	0&	0&	0&	 0 \\	
		-u_x&	1&	0&	0&	0&	0&	0&	0&	0&	0&	0&	0&	0&	0&	0&	0&	0&	0&	0&	0&	0&	0&	0&	0&	0&	0&	 0 \\	
		-u_y&	0&	1&	0&	0&	0&	0&	0&	0&	0&	0&	0&	0&	0&	0&	0&	0&	0&	0&	0&	0&	0&	0&	0&	0&	0&	 0 \\	
		-u_z	&0	&0	&1	&0	&0	&0	&0	&0	&0	&0	&0	&0	&0	&0	&0	&0	&0	&0	&0	&0	&0	&0	&0	&0	&0	& 0 \\	
		u_x u_y &-u_y &-u_x	&0	&1	&0	&0	&0	&0	&0	&0	&0	&0	&0	&0	&0	&0	&0	&0	&0	&0	&0	&0	&0	&0	&0	& 0 \\	
		u_x u_z &-u_z	&0 &-u_x	&0	&1	&0	&0	&0	&0	&0	&0	&0	&0	&0	&0	&0	&0	&0	&0	&0	&0	&0	&0	&0	&0	& 0 \\	
		u_y u_z	&0 &-u_z &-u_y	&0	&0	&1	&0	&0	&0	&0	&0	&0	&0	&0	&0	&0	&0	&0	&0	&0	&0	&0	&0	&0	&0	& 0 \\	
		u_x^2-u_y^2 &-2u_x &2u_y	&0	&0	&0	&0	&1	&0	&0	&0	&0	&0	&0	&0	&0	&0	&0	&0	&0	&0	&0	&0	&0	&0	&0	& 0 \\	
		u_x^2-u_z^2 &-2u_x	&0 &2u_z	&0	&0	&0	&0	&1	&0	&0	&0	&0	&0	&0	&0	&0	&0	&0	&0	&0	&0	&0	&0	&0	&0	& 0 \\	
		u_x^2+u_y^2+u_z^2 &-2u_x &-2u_y &-2u_z	&0	&0	&0	&0	&0	&1	&0	&0	&0	&0	&0	&0	&0	&0	&0	&0	&0	&0	&0	&0	&0	&0	& 0 \\	
		-u_x (u_y^2+ u_z^2) &u_y^2+u_z^2 &2u_x u_y &2u_x u_z &-2u_y &-2u_z	&0 &u_x/3 &u_x/3 &-2u_x/3	&1	&0	&0	&0	&0	&0	&0	&0	&0	&0	&0	&0	&0	&0	&0	&0	& 0 \\	
		-u_y (u_x^2+ u_z^2) &2u_x u_y &u_x^2+u_z^2 &2u_y u_z &-2u_x	&0 &-2u_z &-2u_y/3 &u_y/3 &-2u_y/3	&0	&1	&0	&0	&0	&0	&0	&0	&0	&0	&0	&0	&0	&0	&0	&0	& 0 \\	
		-u_z (u_x^2+ u_y^2) &2u_x u_z &2u_y u_z &u_x^2+u_y^2 &0 &-2u_x &-2u_y &u_z/3 &-2u_z/3 &-2u_z/3	&0	&0	&1	&0	&0	&0	&0	&0	&0	&0	&0	&0	&0	&0	&0	&0	& 0 \\	
		u_x (u_z^2- u_y^2)	 &u_y^2-u_z^2 &2u_x u_y &-2u_x u_z &-2u_y &2u_z	&0 &u_x &-u_x &0	&0	&0	&0	&1	&0	&0	&0	&0	&0	&0	&0	&0	&0	&0	&0	&0	& 0 \\	
		-u_y(u_x^2-u_z^2) &2u_x u_y &u_x^2-u_z^2 &-2u_y u_z &-2ux	&0 &2u_z	&0	&-u_y	&0	&0	&0	&0	&0	&1	&0	&0	&0	&0	&0	&0	&0	&0	&0	&0	&0	& 0 \\	
		-u_z(u_x^2-u_y^2) &2u_x u_z &-2u_y u_z &u_x^2-u_y^2	&0 &-2u_x &2u_y &-u_z &0	&0	&0	&0	&0	&0	&0	&1	&0	&0	&0	&0	&0	&0	&0	&0	&0	&0	& 0 \\	
		-u_x u_y u_z &u_y u_z &u_x u_z &u_x u_y &-u_z &-u_y &-u_x	&0	&0	&0	&0	&0	&0	&0	&0	&0	&1	&0	&0	&0	&0	&0	&0	&0	&0	&0	& 0 \\	
		u_x^2 u_y^2+u_x^2 u_z^2+u_y^2 u_z^2	 &-2u_x (u_y^2+u_z^2) &-2u_y (u_x^2+u_z^2) &-2u_z (u_x^2+u_y^2) &4u_x u_y &4u_x u_z &4u_y u_z &(-u_x^2+2u_y^2-u_z^2)/3 & (-u_x^2-u_y^2+2u_z^2)/3 &2(u_x^2+u_y^2+u_z^2)/3 &-2u_x &-2u_y &-2u_z	&0	&0	&0	&0	&1	&0	&0	&0	&0	&0	&0	&0	&0	& 0 \\	
		u_x^2 u_y^2+u_x^2 u_z^2-u_y^2 u_z^2	 &-2u_x (u_y^2+u_z^2) &-2u_y (u_x^2-u_z^2) &-2u_z (u_x^2-u_y^2) &4u_x u_y &4u_x u_z &-4u_y u_z &u_z^2-u_x^2/3 &u_y^2-u_x^2/3 &2u_x^2/3 &-2u_x	&0	&0	&0 &-2u_y &-2u_z	&0	&0	&1	&0	&0	&0	&0	&0	&0	&0	& 0 \\	
		u_x^2 (u_y^2- u_z^2) &2u_x (u_z^2-u_y^2) &-2 u_x^2 u_y &2u_x^2 u_z &4u_x u_y &-4u_x u_z	&0 &(-3u_x^2+u_y^2-u_z^2)/3 &(3u_x^2+u_y^2-u_z^2)/3 &(u_y^2-u_z^2)/3	&0 &-u_y &u_z &-2u_x &-u_y &u_z	&0	&0	&0	&1	&0	&0	&0	&0	&0	&0	& 0 \\	
		u_x^2 u_y u_z &-2 u_x u_y u_z &-u_x^2 u_z &-u_x^2 u_y &2u_x u_z &2u_x u_y &u_x^2 &u_y u_z/3 &u_y u_z/3 &u_y u_z/3	&0 &-u_z/2 &-u_y/2	&0 &-u_z/2 &-u_y/2 &-2u_x	&0	&0	&0	&1	&0	&0	&0	&0	&0	& 0 \\	
		u_x u_y^2 u_z &-u_y^2 u_z &-2u_x u_y u_z &-u_x u_y^2 &2u_y u_z &u_y^2 &2u_x u_y &-2u_x u_z/3 &u_x u_z/3 &u_x u_z/3 &-u_z/2	&0 &-u_x/2 &-u_z/2	&0 &u_x/2 &-2u_y	&0	&0	&0	&0	&1	&0	&0	&0	&0	& 0 \\	
		u_x u_y u_z^2 &-u_y u_z^2 &-u_x u_z^2 &-2u_x u_y u_z &u_z^2 &2u_y u_z &2u_x u_z &u_x u_y/3 &-2u_x u_y/3 &u_x u_y/3 &-u_y/2 &-u_x/2	&0 &u_y/2 &u_x/2	&0 &-2u_z	&0	&0	&0	&0	&0	&1	&0	&0	&0	& 0 \\	
		-u_x u_y^2 u_z^2 &u_y^2 u_z^2 &2u_x u_y u_z^2 &2u_x u_y^2 u_z &-2 u_y u_z^2 &-2u_y^2 u_z &-4u_x u_y u_z &u_x(2u_z^2-u_y^2)/3 &u_x(2u_y^2-u_z^2)/3 &u_x(-u_y^2-u_z^2)/3 &(u_y^2+u_z^2)/2 &u_x u_y &u_x u_z &(u_z^2 - u_y^2)/2 &-u_x u_y &-u_x u_z &4u_y u_z &-u_x/2 &u_x/2	&0	&0 &-2u_z &-2u_y	&1	&0	&0	& 0 \\	
		-u_x^2 u_y u_z^2 &2u_x u_y u_z^2 &u_x^2 u_z^2 &2u_x^2 u_y u_z &-2u_x u_z^2 &-4u_x u_y u_z &-2u_x^2 u_z &u_y(-u_x^2-u_z^2)/3 &u_y(2u_x^2-u_z^2)/3 &u_y(-u_x^2-u_z^2)/3 &u_x u_y &(u_x^2+u_z^2)/2 &u_y u_z &-u_x u_y &(u_z^2-u_x^2)/2 &u_y u_z &4u_x u_z &-u_y/4 &-u_y/4 &u_y/2 &-2u_z	&0 &-2u_x	&0	&1	&0	& 0 \\	
		-u_x^2 u_y^2 u_z &2u_x u_y^2 u_z &2u_x^2 u_y u_z &u_x^2 u_y^2 &-4u_x u_y u_z &-2u_x u_y^2 &-2u_x^2 u_y &u_z(2u_x^2-u_y^2)/3 &u_z(-u_x^2-u_y^2)/3 &u_z(-u_x^2-u_y^2)/3 &u_x u_z &u_y u_z &(u_x^2+u_y^2)/2 &u_x u_z &u_y u_z &(u_y^2-u_x^2)/2 &4u_x u_y &-u_z/4 &-u_z/4 &-u_z/2 &-2u_y &-2u_x &0	&0	&0	&1	& 0 \\	
		u_x^2 u_y^2 u_z^2 &-2u_x u_y^2 u_z^2 &-2u_x^2 u_y u_z^2 &-2u_x^2 u_y^2 u_z &4u_x u_y u_z^2 &4u_x u_y^2 u_z &4u_x^2 u_y u_z &(u_x^2 u_y^2-2u_x^2u_z^2+u_y^2u_z^2)/3	 &(-2u_x^2 u_y^2+ u_x^2 u_z^2+u_y^2 u_z^2)/3	 &(u_x^2 u_y^2+u_x^2 u_z^2+u_y^2 u_z^2)/3	 &-u_x (u_y^2+u_z^2)	&-u_y (u_x^2+u_z^2) &-u_z (u_x^2+u_y^2)	 &u_x (u_y^2-u_z^2) &u_y (u_x^2-u_z^2)	 &u_z (u_x^2-u_y^2)	 &-8u_x u_y u_z &(2u_x^2+u_y^2+u_z^2)/4 &(-2u_x^2+u_y^2+u_z^2)/4 &(u_z^2-u_y^2)/2	&4u_y u_z &4u_x u_z &4u_x u_y &-2u_x &-2u_y &-2u_z &1 \\	
	\end{array}
	\right] $},
	}
\end{align}
and its inverse ${\bf N}^{-1}$ can be given by
\begin{equation}
\thickmuskip=0mu
\medmuskip=0mu
\thinmuskip=0mu
\rotatebox{90}{
\scalebox{0.45}{$\displaystyle
{\bf N}^{-1} = 
\left [
	\begin{array}{ccccccccccccccccccccccccccc}
		1	&0&	0&	0&	0&	0&	0&	0&	0&	0&	0&	0&	0&	0&	0&	0&	0&	0&	0&	0&	0&	0&	0&	0&	0&	0&	 0 \\	
		u_x&                 1&                 0&                 0&         0&         0&         0&                                         0&                                           0&                                       0&             0&             0&             0&             0&               0&               0&       0&                     0&                       0&             0&     0&     0&     0&   0&   0&   0& 0 \\	
		u_y&                 0&                 1&                 0&         0&         0&         0&                                         0&                                           0&                                       0&             0&             0&             0&             0&               0&               0&       0&                     0&                       0&             0&     0&     0&     0&   0&   0&   0& 0 \\	
		u_z&                 0&                 0&                 1&         0&         0&         0&                                         0&                                           0&                                       0&             0&             0&             0&             0&               0&               0&       0&                     0&                       0&             0&     0&     0&     0&   0&   0&   0& 0 \\	
		u_xu_y&                 u_y&                 u_x&                 0&         1&         0&         0&                                         0&                                           0&                                       0&             0&             0&             0&             0&               0&               0&       0&                     0&                       0&             0&     0&     0&     0&   0&   0&   0& 0 \\	
		u_xu_z&                 u_z&                 0&                 u_x&         0&         1&         0&                                         0&                                           0&                                       0&             0&             0&             0&             0&               0&               0&       0&                     0&                       0&             0&     0&     0&     0&   0&   0&   0& 0 \\	
		u_yu_z&                 0&                 u_z&                 u_y&         0&         0&         1&                                         0&                                           0&                                       0&             0&             0&             0&             0&               0&               0&       0&                     0&                       0&             0&     0&     0&     0&   0&   0&   0& 0 \\	
		u_x^2 - u_y^2&               2u_x&              -2u_y&                 0&         0&         0&         0&                                         1&                                           0&                                       0&             0&             0&             0&             0&               0&               0&       0&                     0&                       0&             0&     0&     0&     0&   0&   0&   0& 0 \\	
		u_x^2 - u_z^2&               2u_x&                 0&              -2u_z&         0&         0&         0&                                         0&                                           1&                                       0&             0&             0&             0&             0&               0&               0&       0&                     0&                       0&             0&     0&     0&     0&   0&   0&   0& 0 \\	
		u_x^2 + u_y^2 + u_z^2&               2u_x&               2u_y&               2u_z&         0&         0&         0&                                         0&                                           0&                                       1&             0&             0&             0&             0&               0&               0&       0&                     0&                       0&             0&     0&     0&     0&   0&   0&   0& 0 \\	
		u_x(u_y^2 + u_z^2)&         u_y^2 + u_z^2&             2u_xu_y&             2u_xu_z&       2u_y&       2u_z&         0&                                      -u_x/3&                                        -u_x/3&                                 2u_x/3&             1&             0&             0&             0&               0&               0&       0&                     0&                       0&             0&     0&     0&     0&   0&   0&   0& 0 \\	
		u_y(u_x^2 + u_z^2)&             2u_xu_y&         u_x^2 + u_z^2&             2u_yu_z&       2u_x&         0&       2u_z&                                   2u_y/3&                                        -u_y/3&                                 2u_y/3&             0&             1&             0&             0&               0&               0&       0&                     0&                       0&             0&     0&     0&     0&   0&   0&   0& 0 \\	
		u_z(u_x^2 + u_y^2)&             2u_xu_z&             2u_yu_z&         u_x^2 + u_y^2&         0&       2u_x&       2u_y&                                      -u_z/3&                                     2u_z/3&                                 2u_z/3&             0&             0&             1&             0&               0&               0&       0&                     0&                       0&             0&     0&     0&     0&   0&   0&   0& 0 \\	
		u_x(u_y^2 - u_z^2)&         u_y^2 - u_z^2&             2u_xu_y&            -2u_xu_z&       2u_y&      -2u_z&         0&                                        -u_x&                                           u_x&                                       0&             0&             0&             0&             1&               0&               0&       0&                     0&                       0&             0&     0&     0&     0&   0&   0&   0& 0 \\	
		u_y(u_x^2 - u_z^2)&             2u_xu_y&         u_x^2 - u_z^2&            -2u_yu_z&       2u_x&         0&      -2u_z&                                         0&                                           u_y&                                       0&             0&             0&             0&             0&               1&               0&       0&                     0&                       0&             0&     0&     0&     0&   0&   0&   0& 0 \\	
		u_z(u_x^2 - u_y^2)&             2u_xu_z&            -2u_yu_z&         u_x^2 - u_y^2&         0&       2u_x&      -2u_y&                                         u_z&                                           0&                                       0&             0&             0&             0&             0&               0&               1&       0&                     0&                       0&             0&     0&     0&     0&   0&   0&   0& 0 \\	
		u_xu_yu_z&               u_yu_z&               u_xu_z&               u_xu_y&         u_z&         u_y&         u_x&                                         0&                                           0&                                       0&             0&             0&             0&             0&               0&               0&       1&                     0&                       0&             0&     0&     0&     0&   0&   0&   0& 0 \\	
		u_x^2u_y^2 + u_x^2u_z^2 + u_y^2u_z^2& 2u_x(u_y^2 + u_z^2)& 2u_y(u_x^2 + u_z^2)& 2u_z(u_x^2 + u_y^2)&     4u_xu_y&     4u_xu_z&     4u_yu_z&               (- u_x^2 + 2u_y^2 - u_z^2)/3&                 (- u_x^2 - u_y^2 + 2u_z^2)/3&       2(u_x^2 + u_y^2 + u_z^2)/3&           2u_x&           2u_y&           2u_z&             0&               0&               0&       0&                     1&                       0&             0&     0&     0&     0&   0&   0&   0& 0 \\	
		u_x^2u_y^2 + u_x^2u_z^2 - u_y^2u_z^2& 2u_x(u_y^2 + u_z^2)& 2u_y(u_x^2 - u_z^2)& 2u_z(u_x^2 - u_y^2)&     4u_xu_y&     4u_xu_z&    -4u_yu_z&                               u_z^2 - u_x^2/3&                                 u_y^2 - u_x^2/3&                               2u_x^2/3&           2u_x&             0&             0&             0&             2u_y&             2u_z&       0&                     0&                       1&             0&     0&     0&     0&   0&   0&   0& 0 \\	
		u_x^2(u_y^2 - u_z^2)& 2u_x(u_y^2 - u_z^2)&           2u_x^2u_y&          -2u_x^2u_z&     4u_xu_y&    -4u_xu_z&         0&                     (- 3u_x^2 + u_y^2 - u_z^2)/3&                         (3u_x^2 + u_y^2 - u_z^2)/3&                           (u_y^2 - u_z^2)/3&             0&             u_y&            -u_z&           2u_x&               u_y&              -u_z&       0&                     0&                       0&             1&     0&     0&     0&   0&   0&   0& 0 \\	
		u_x^2u_yu_z&           2u_xu_yu_z&             u_x^2u_z&             u_x^2u_y&     2u_xu_z&     2u_xu_y&       u_x^2&                                   u_yu_z/3&                                     u_yu_z/3&                                 u_yu_z/3&             0&           u_z/2&           u_y/2&             0&             u_z/2&             u_y/2&     2u_x&                     0&                       0&             0&     1&     0&     0&   0&   0&   0& 0 \\	
		u_xu_y^2u_z&             u_y^2u_z&           2u_xu_yu_z&             u_xu_y^2&     2u_yu_z&       u_y^2&     2u_xu_y&                                -2u_xu_z/3&                                     u_xu_z/3&                                 u_xu_z/3&           u_z/2&             0&           u_x/2&           u_z/2&               0&            -u_x/2&     2u_y&                     0&                       0&             0&     0&     1&     0&   0&   0&   0& 0 \\	
		u_xu_yu_z^2&             u_yu_z^2&             u_xu_z^2&           2u_xu_yu_z&       u_z^2&     2u_yu_z&     2u_xu_z&                                   u_xu_y/3&                                  -2u_xu_y/3&                                 u_xu_y/3&           u_y/2&           u_x/2&             0&          -u_y/2&            -u_x/2&               0&     2u_z&                     0&                       0&             0&     0&     0&     1&   0&   0&   0& 0 \\	
		u_xu_y^2u_z^2&           u_y^2u_z^2&         2u_xu_yu_z^2&         2u_xu_y^2u_z&   2u_yu_z^2&   2u_y^2u_z&   4u_xu_yu_z&                   u_x(u_y^2 - 2u_z^2)/3&                     u_x(u_z^2 - 2u_y^2)/3&                   u_x(u_y^2 + u_z^2)/3& (u_y^2 + u_z^2)/2&           u_xu_y&           u_xu_z& (u_z^2 - u_y^2)/2&            -u_xu_y&            -u_xu_z&   4u_yu_z&                   u_x/2&                    -u_x/2&             0&     0&   2u_z&   2u_y&   1&   0&   0& 0 \\	
		u_x^2u_yu_z^2&         2u_xu_yu_z^2&           u_x^2u_z^2&         2u_x^2u_yu_z&   2u_xu_z^2&   4u_xu_yu_z&   2u_x^2u_z&                     u_y(u_x^2 + u_z^2)/3&                     u_y(u_z^2 - 2u_x^2)/3&                   u_y(u_x^2 + u_z^2)/3&           u_xu_y& (u_x^2 + u_z^2)/2&           u_yu_z&          -u_xu_y&   (u_z^2 - u_x^2)/2&             u_yu_z&   4u_xu_z&                   u_y/4&                     u_y/4&          -u_y/2&   2u_z&     0&   2u_x&   0&   1&   0& 0 \\	
		u_x^2u_y^2u_z&         2u_xu_y^2u_z&         2u_x^2u_yu_z&           u_x^2u_y^2&   4u_xu_yu_z&   2u_xu_y^2&   2u_x^2u_y&                   u_z(u_y^2 - 2u_x^2)/3&                       u_z(u_x^2 + u_y^2)/3&                   u_z(u_x^2 + u_y^2)/3&           u_xu_z&           u_yu_z& (u_x^2 + u_y^2)/2&           u_xu_z&             u_yu_z&   (u_y^2 - u_x^2)/2&   4u_xu_y&                   u_z/4&                     u_z/4&           u_z/2&   2u_y&   2u_x&     0&   0&   0&   1& 0 \\	
		u_x^2u_y^2u_z^2&       2u_xu_y^2u_z^2&       2u_x^2u_yu_z^2&       2u_x^2u_y^2u_z& 4u_xu_yu_z^2& 4u_xu_y^2u_z& 4u_x^2u_yu_z& (u_x^2u_y^2 - 2u_x^2u_z^2 + u_y^2u_z^2)/3& (-2u_x^2u_y^2 + u_x^2u_z^2 + u_y^2u_z^2)/3& (u_x^2u_y^2 + u_x^2u_z^2 + u_y^2u_z^2)/3& u_x(u_y^2 + u_z^2)& u_y(u_x^2 + u_z^2)& u_z(u_x^2 + u_y^2)& u_x(u_z^2 - u_y^2)& u_y(u_z^2 - u_x^2)& u_z(u_y^2 - u_x^2)& 8u_xu_yu_z& (2u_x^2 + u_y^2 + u_z^2)/4& (- 2u_x^2 + u_y^2 + u_z^2)/4& (u_z^2 - u_y^2)/2& 4u_yu_z& 4u_xu_z& 4u_xu_y& 2u_x& 2u_y& 2u_z& 1 \\	
	\end{array}
	\right] $}.
	} \label{eq:invN}
\end{equation}
Note that the above practical forms [Eqs.~(\ref{eq:M})--(\ref{eq:invN})] depend on the definition of the lattice velocity ${\bf c}_i$.
\end{widetext}


\bibliography{refs}

\end{document}